\documentclass[twocolumn]{aastex61}

\newcommand\aastex{AAS\TeX}

\received{}
\revised{}
\accepted{}
\submitjournal{ApJ}

\shorttitle{\aastex\ Probing the structure of \emph{Kepler} ZZ Ceti stars}
\shortauthors{Romero et al.}

\begin{document}

\title{Probing the structure of \emph{Kepler} ZZ Ceti stars with full
  evolutionary models-based asteroseismology}

\correspondingauthor{Alejandra Romero}
\email{alejandra.romero@ufrgs.br}

\author{Alejandra D. Romero}
\affiliation{Departamento de Astronomia, Universidade Federal do Rio 
          Grande do Sul, Av. Bento Goncalves 9500, Porto Alegre 91501-970, 
          RS, Brazil}
\nocollaboration

\author{A. H. C\'orsico}
\affiliation{Grupo de Evoluci\'on Estelar y Pulsaciones, 
          Facultad de Ciencias Astron\'omicas y Geof\'{\i}sicas, 
          Universidad Nacional de La Plata, Paseo del Bosque s/n, 1900 
          La Plata, Argentina}
\affiliation{Instituto de Astrof\'isica La Plata, CONICET-UNLP, Argentina}
\nocollaboration

\author{B. G. Castanheira}
\affiliation{Department of Astronomy and McDonald Observatory, University of Texas at Austin, Austin, TX 78712, USA}

\author{F. C. De Ger\'onimo}
\affiliation{Grupo de Evoluci\'on Estelar y Pulsaciones, 
          Facultad de Ciencias Astron\'omicas y Geof\'{\i}sicas, 
          Universidad Nacional de La Plata, Paseo del Bosque s/n, 1900 
          La Plata, Argentina}
\affiliation{Instituto de Astrof\'isica La Plata, CONICET-UNLP, Argentina}

\author{S. O. Kepler}
\affiliation{Departamento de Astronomia, Universidade Federal do Rio 
          Grande do Sul, Av. Bento Goncalves 9500, Porto Alegre 91501-970, 
          RS, Brazil}

\author{D. Koester}
\affiliation{Institut f\"ur Theoretische Physik und Astrophysik, Universit\"at Kiel, 24098 Kiel, Germany}

\author{A. Kawka}
\affiliation{Astronomick\'y \'ustav, Akademie v\u ed \u Cesk\'e republiky, Fri\u cova 298, CZ-251 65 Ond\u rejov, Czech Republic}

\author{L. G. Althaus}
\affiliation{Grupo de Evoluci\'on Estelar y Pulsaciones, 
          Facultad de Ciencias Astron\'omicas y Geof\'{\i}sicas, 
          Universidad Nacional de La Plata, Paseo del Bosque s/n, 1900 
          La Plata, Argentina}
\affiliation{Instituto de Astrof\'isica La Plata, CONICET-UNLP, Argentina}

\author{J. J. Hermes}
\affiliation{Department of Physics and Astronomy, University of North Carolina, Chapel Hill, NC 27599-3255, USA}

\author{C. Bonato}
\affiliation{Departamento de Astronomia, Universidade Federal do Rio 
          Grande do Sul, Av. Bento Goncalves 9500, Porto Alegre 91501-970, 
          RS, Brazil}

\author{A. Gianninas}
\affiliation{Homer L. Dodge Department of Physics and Astronomy, University of Oklahoma, 440 W. Brooks St., Norman, OK 73019, USA}



\begin{abstract}
 We present  an asteroseismological analysis  of four  ZZ Ceti stars
 observed with \emph{Kepler}: GD 1212, SDSS J113655.17+040952.6, KIC
 11911480  and KIC 4552982, based  on a grid of  full evolutionary
 models   of  DA   white  dwarf   stars.  We   employ a grid of
 carbon-oxygen core white  dwarfs models, characterized  by a detailed
 and consistent  chemical inner profile for the  core and the
 envelope. In addition  to the observed periods, we  take into account
 other  information  from  the   observational  data,  as  amplitudes,
 rotational splittings and  period spacing, as well  as photometry and
 spectroscopy. For  each star,  we present  an asteroseismological
 model that closely reproduce their observed  properties.  The
 asteroseismological  stellar mass  and effective  temperature of the
 target stars are ($0.632  \pm 0.027  M_{\sun}$, $10737  \pm 73$K)
 for GD 1212, ($0.745 \pm  0.007 M_{\sun}$, $11110 \pm 69$K) for KIC
 4552982, ($0.5480  \pm  0.01 M_{\sun}$,  $12721  \pm 228$K)  for
 KIC1191480 and ($0.570 \pm 0.01 M_{\sun}$, $12060 \pm 300$K) for
 SDSS J113655.17+040952.6.   In general, the  asteroseismological
 values  are  in  good agreement with the spectroscopy.   
 For KIC 11911480  and SDSS J113655.17+040952.6  we
 derive a  similar seismological  mass,  but  the hydrogen  envelope
 is   an  order  of magnitude   thinner  for  SDSS
 J113655.17+040952.6, that is part of a binary system and went through
 a common envelope phase.

\end{abstract}




\section{Introduction} 

\label{introduction}

ZZ Ceti (or DAV) variable  stars constitute the most populous class of
pulsating  white   dwarfs (WDs).  They  are  otherwise   normal  DA
(H-rich atmospheres) WDs located in a narrow  instability strip with
effective   temperatures   between  $10\,500$   K   and  $12\,500$   K
\citep[e.g.,][]{2008ARA&A..46..157W,2008PASP..120.1043F, review,
  2017EPJWC.15201011K}    that show luminosity  variations of  up to
$0.30$  mag caused  by nonradial $g$-mode pulsations of low degree
($\ell \leq 2$) and periods between 70  and 1500  s.  Pulsations  are
triggered  by a  combination  of the $\kappa-\gamma$ mechanism acting
at  the basis of the hydrogen partial ionization
zone
\citep{1981A&A...102..375D,1981A&A....97...16D,1982ApJ...252L..65W}
and     the    convective     driving     mechanism
\citep{1991MNRAS.252..334B,1999ApJ...511..904G}.

Asteroseismology  of WDs  uses  the comparison  of the observed
pulsation periods  with the  adiabatic periods  computed for
appropriate stellar  models. It allows  us to learn about  the origin,
internal     structure    and     evolution     of    WDs
\citep{2008ARA&A..46..157W,review,2008PASP..120.1043F}.  In
particular, asteroseismological   analysis  of  ZZ   Ceti  stars
provide  strong constraints on the stellar mass, the thickness of the
outer envelopes, the  core  chemical  composition,  and  the  stellar
rotation  rates.  Furthermore,  the rate of  period changes  of ZZ
Ceti stars  allows to derive               the               cooling
timescale
\citep{2005ApJ...634.1311K,2012ASPC..462..322K,2013ApJ...771...17M},
to                             study                            axions
\citep{Isern92,2001NewA....6..197C,2008ApJ...675.1512B,2012MNRAS.424.2792C,2012JCAP...12..010C,
  2016JCAP...07..036C}, neutrinos
\citep{2004ApJ...602L.109W,2014JCAP...08..054C},  and  the possible
secular  rate of  variation  of  the gravitational  constant
\citep{2013JCAP...06..032C}.  Finally, ZZ  Ceti stars  allow  to study
crystallization
\citep{1999ApJ...526..976M,2004A&A...427..923C,2005A&A...429..277C,2004ApJ...605L.133M,2005A&A...432..219K,2013ApJ...779...58R},
to          constrain          nuclear         reaction          rates
\citep[e.g.
  $^{12}$C$(\alpha,\gamma)^{16}$O,][]{2002ApJ...573..803M}, to   infer
the   properties    of   the   outer   convection   zones
\citep{2005ApJ...633.1142M,2005ASPC..334..483M,2007ASPC..372..635M},
and   to   look   for   extra-solar   planets   orbiting   these
stars \citep{2008ApJ...676..573M}.

Two  main  approaches  have  been  adopted hitherto  for  WD
asteroseismology.  One  of them  employs  stellar models  with
parametrized chemical profiles. This approach has the advantage that
it allows a full exploration of parameter space to find the best
seismic model \citep[see, for
  details,][]{2011ApJ...742L..16B,2014ApJ...794...39B,2016ApJS..223...10G,2017ApJ...834..136G,2017A&A...598A.109G}.
However, this method requires the number of detected periods to be larger to the number of free parameters in the model grid, which is not always the case for pulsationg DA stars. 
The other approach ---the  one we
adopt in this paper--- employs  fully   evolutionary  models
resulting   from  the  complete evolution of  the progenitor stars,
from  the ZAMS to  the WD stage.  Because this approach is more time
consuming from the computational point of view, usually the model
grid is not as thin or versatile as in the first approach. However, it
involves the  most detailed  and  updated input physics, in particular
regarding  the internal chemical structure from the  stellar  core  to
the  surface,  that  is  a crucial  aspect  for correctly
disentangling  the   information  encoded  in  the  pulsation patterns
of variable WDs. Specially, most structural parameters are set
consistently by the evolution prior to the white dwarf cooling phase,
reducing significantly the number of free parameters. The use of full
evolutionary models has been extensively applied in
asteroseismological  analysis of hot GW Vir  (or DOV) stars
\citep{2007A&A...461.1095C,2007A&A...475..619C,2008A&A...478..869C,2009A&A...499..257C,2014MNRAS.442.2278K,
  2016arXiv160206355C}, V777                  Her
(DBV)                 stars
\citep{2012A&A...541A..42C,2014A&A...570A.116B,2014JCAP...08..054C},
ZZ                   Ceti                  stars
\citep{2012ApJ...757..177K,2012MNRAS.420.1462R,2013ApJ...779...58R},
and Extremely low mass white dwarf variable stars (ELMV)\footnote{Extremely low mass white dwarf stars are He-core white dwarf stars with stellar masses below $\sim 0.3 M_{\odot}$ \citep{2010ApJ...723.1072B}) and are thought to be the result of strong-mass transfer events in close binary systems.} \citep{2017arXiv170800482C}. 

Out   of   the   $\sim$170    ZZ   Ceti   stars   known   to   date
\citep{2016IBVS.6184....1B,2017EPJWC.15201011K}\footnote{Not including the  recently
  discovered pulsating low mass- and extremely low-mass WDs
  \citep[]{2012ApJ...750L..28H,2013ApJ...765..102H,2013MNRAS.436.3573H,2015MNRAS.446L..26K,2016arXiv161206390B}.},
48 are bright  objects with $V<16$,  and the remainder  are fainter
ZZ Ceti stars that have been detected with the Sloan Digital Sky
Survey (SDSS) \citep{2004ApJ...607..982M,2005ApJ...625..966M,
  2005A&A...442..629K,2012ApJ...757..177K,2006A&A...450..227C,2007A&A...462..989C,2010MNRAS.405.2561C,2013MNRAS.430...50C}. The
list  is now being enlarged with  the recent  discovery of  pulsating
WD  stars within the  \emph{Kepler} spacecraft field, thus opening a
new avenue for WD asteroseismology based on observations from
space \citep[see e.g.][]{2017arXiv170907004H}. 
This kind of data is different from ground base photometry
  because it does not have the usual gaps due to daylight, but also
  different reduction techniques have to be employed to uncover the
  pulsation spectra of the stars observed with the {\emph Kepler}
  spacecraft.  In particular, after the two wheels stopped to
  function, known as the K2 phase, additional noise is introduced to
  the signal due to the shooting of the trusters with a timescale around six hours to
  correct the pointing. The  ZZ  Ceti  longest observed by
\emph{Kepler}, KIC 4552982 (WD J1916+3938, $T_{\rm eff}=  10\, 860$ K,
$\log g = 8.16$), was      discovered      from      ground-based
photometry      by \citet{2011ApJ...741L..16H}\footnote{Almost
  simultaneously, the first DBV  star in  the \emph{Kepler}  Mission
  field,  KIC  8626021 (GALEX J1910+4425), was  discovered by
  \citet{2011ApJ...736L..39O}.}.  This star exhibits  pulsation
periods in  the range $360-1500$ s  and shows energetic outbursts
\citep[][]{2015ApJ...809...14B}.  A  second ZZ Ceti  star observed
with  \emph{Kepler}  is  KIC 11911480  (WD  J1920+5017,  $T_{\rm
  eff}= 12\,160$ K, $\log g = 7.94$), that exhibits a total of six
independent pulsation modes with periods between 173 and 325 s
\citep{2014MNRAS.438.3086G},  typical of the hot ZZ Ceti stars
\citep{2000MNRAS.314..220C,2006ApJ...640..956M}. Four of its pulsation
modes show strong signatures of rotational splitting, allowing to
estimate a rotation period of  $\sim$3.5 days. The  ZZ Ceti star GD
1212 (WD  J2338$-$0741, $T_{\rm  eff}= 10\, 980$  K,   $\log g  =
7.995$, \citep{2017arXiv170907004H}  was observed for a total of 264.5 hr using the \emph{Kepler}
(K2) spacecraft in two-wheel mode. \citep{2014ApJ...789...85H}
reported the detection of 19 pulsation modes, with  periods ranging
from 828 to 1221 s. Recently \citet{2017arXiv170907004H} analyzed the light curve and find a 
smaller number of real $m=0$ component modes in the spectra, which we will consider to performe our seismological analysis. Finally,
there is the ZZ Ceti star SDSS J113655.17+040952.6 (J1136$+$0409),
discovered by  \citet{2015MNRAS.447..691P} and observed in detail  by
\citet{2015MNRAS.451.1701H}. This is the first known DAV variable  WD
in a post--common--envelope binary system. Recently,
\citet{2016MNRAS.457.2855G} reported additional ZZ Ceti stars in the
\emph{Kepler} mission field. Also, \citet{2017arXiv170907004H}  
present photometry and spectroscopy for 27 ZZ Ceti stars observed by
the \emph{ Kepler} space telescope, including the four objects analyzed here.

In this paper, we carry out an asteroseismological analysis of
the first four published ZZ Ceti stars observed with \emph{Kepler} by
employing evolutionary DA WD models representative of these objects.
We  perform our  study by employing  a grid  of full  evolutionary
models  representative  of DA WD stars as  discussed  in
\citet{2012MNRAS.420.1462R}  and extended     toward     higher
stellar     mass     values     in
\citet{2013ApJ...779...58R}. Evolutionary models have consistent
chemical profiles for  both  the core  and  the  envelope  for various
stellar  masses, specifically calculated for asteroseismological fits
of ZZ Ceti stars.  The chemical profiles of our models are computed
considering the complete evolution  of the progenitor stars from  the
ZAMS through the  thermally pulsing and  mass-loss phases  on the
asymptotic giant branch  (AGB).  Our asteroseismological approach
combines (1) a significant exploration of the  parameter  space $(M_{\star},
T_{\rm  eff},  M_{\rm  H})$, and  (2) updated input  physics,  in
particular, regarding  the internal  chemical structure,  that  is   a
crucial  aspect  for  WD asteroseismology. In addition, the impact of
the uncertainties resulting from the evolutionary history of
progenitor star on the properties of asteroseismological models of ZZ
Ceti stars  has been assessed by \cite{2017A&A...599A..21D} and
De Ger\'onimo et al. (2017b, submitted.). This
adds confidence to the use of fully evolutionary models with consistent
chemical profiles, and renders much more robust our asteroseismological
approach.

The  paper is  organized as  follows.  In  Sect.   \ref{numerical}, we
provide  a  brief description  of  the  evolutionary  code, the  input
physics adopted in  our calculations and the grid  of models employed.
In Sect.  \ref{astero},  we describe our asteroseismological procedure
and  the  application  to  the  target stars.  We  conclude  in  Sect.
\ref{conclusions} by summarizing our findings.

\section{Numerical tools and models}
\label{numerical}

\subsection{Input physics}
\label{input}

The grid of full evolutionary models used in this work was calculated
with  an  updated  version  of  the  {\tt  LPCODE}  evolutionary  code
\citep[see][for
  details]{2005A&A...435..631A,2010ApJ...717..897A,2010ApJ...717..183R,2015MNRAS.450.3708R}. {\tt
  LPCODE} compute the evolution of single, i.e. non--binary, stars with low
and intermediate mass at the Main Sequence.  Here, we briefly  mention
the main input  physics relevant for this work.  Further  details can
be  found in  those  papers  and in
\citet{2012MNRAS.420.1462R,2013ApJ...779...58R}.

The {\tt LPCODE} evolutionary  code considers a simultaneous treatment
of    no-instantaneous     mixing    and    burning     of    elements
\citep{2003A&A...404..593A}.  The  nuclear network accounts explicitly
for 16  elements and  34 nuclear reactions,  that include  $pp$ chain,
CNO-cycle,      helium      burning      and      carbon      ignition
\citep{2010ApJ...717..183R}.   In  particular,  the  $^{12}$C$(\alpha,
\gamma)^{16}$O   reaction   rate,  of   special   relevance  for   the
carbon-oxygen  stratification  of the  resulting  WD,  was taken  from
\citet{1999NuPhA.656....3A}.           Note          that          the
$^{12}$C$(\alpha,\gamma)^{16}$O  reaction  rate  is  one of  the  main
source of uncertainties in stellar evolution.  By considering  the
computations of \citet{2002ApJ...567..643K} for    the
$^{12}$C$(\alpha, \gamma)^{16}$O  reaction rate,  the  oxygen
abundance  at the center  can  vary from 26\% to  45\% within  the
theoretical  uncertainties, leading to a change in the period values
up to $\sim 11$ s for a stellar mass of 0.548$M_{\sun}$
\citep{2017A&A...599A..21D}.   We  consider  the  occurrence  of
extra-mixing  episodes  beyond  each convective     boundary
following     the     prescription     of \citet{1997A&A...324L..81H},
except for  the thermally  pulsating AGB phase. We considered mass
loss  during the core helium burning and the red  giant branch  phases
following \citet{2005ApJ...630L..73S},  and during the AGB and
thermally pulsating AGB following the prescription of
\citet{1993ApJ...413..641V}.  During  the WD evolution, we considered
the distinct  physical  processes that  modify the inner chemical
profile. In particular, element  diffusion strongly affects  the
chemical  composition  profile   throughout  the  outer
layers. Indeed, our sequences  develop a pure hydrogen envelope with
increasing  thickness as  evolution  proceeds. Our  treatment of  time
dependent  diffusion  is based  on  the  multicomponent gas  treatment
presented  in \citet{1969fecg.book.....B}.  We  consider gravitational
settling  and thermal  and chemical  diffusion of  H,  $^3$He, $^4$He,
$^{12}$C, $^{13}$C, $^{14}$N and $^{16}$O \citep{2003A&A...404..593A}.
To  account for convection process  we adopted  the mixing length
theory,  in its  ML2 flavor, with  the free  parameter $\alpha= 1.61$
\citep{1990ApJS...72..335T} during the evolution previous to the
white dwarf cooling curve, and $\alpha =1$ during the white dwarf
evolution.    Last, we considered  the chemical rehomogenization  of
the   inner  carbon-oxygen  profile  induced  by Rayleigh-Taylor
instabilities following \citet{1997ApJ...486..413S}.

The  input  physics of  the  code includes  the  equation  of state
of \citet{1994ApJ...434..641S} for  the high density  regime
complemented with   an   updated   version    of   the   equation   of
state   of \citet{1979A&A....72..134M} for the low density
regime. Other physical ingredients  considered in  {\tt LPCODE}  are
the  radiative opacities from the OPAL opacity project
\citep{1996ApJ...464..943I} supplemented at    low    temperatures
with    the    molecular   opacities    of
\citet{1994ApJ...437..879A}.   Conductive  opacities  are  those  from
\citet{2007ApJ...661.1094C}, and the neutrino emission rates are taken
from \citet{1996ApJS..102..411I} and \citet{1994ApJ...425..222H}.

Cool  WD stars  are expected  to crystallize  as a  result of strong
Coulomb   interactions   in   their   very   dense   interior
\citep{1968ApJ...151..227V}. In the process two additional energy
sources, i.e. the release  of latent heat and the release of  gravitational
energy  associated  with changes  in  the  chemical composition  of
carbon-oxygen   profile  induced  by  crystallization
\citep{1988Natur.333..642G,1988A&A...193..141G,2009ApJ...693L...6W}
are considered self-consistently and locally coupled  to the  full set
of  equations of  stellar evolution.  The  chemical redistribution due
to  phase  separation  has  been considered  following  the  procedure
described    in    \citet{1999ApJ...526..976M}   and
\citet{1997ApJ...486..413S}.  To assess the  enhancement of  oxygen in
the  crystallized core we used the azeotropic-type formulation of
\citet{2010PhRvL.104w1101H}.

\subsection{Model grid}
\label{models}

The DA  WD models  used in this  work are the result  of full
evolutionary  calculations of  the  progenitor stars,  from the  ZAMS,
through  the hydrogen  and  helium central  burning stages,  thermal
pulses, the planetary nebula phase and finally the white dwarf
cooling sequences,  using the {\tt  LPCODE} code. The metallicity
value adopted in the main sequence models is $Z=0.01$.   Most of the
sequences with  masses $\lesssim  0.878 M_{\sun}$  were used in the
asteroseismological  study of 44 bright ZZ Ceti stars by
\citet{2012MNRAS.420.1462R},  and  were  extracted from  the  full
evolutionary  computations  of \citet{2010ApJ...717..897A}  \citep[see
  also][]{2010ApJ...717..183R}.   \citet{2013ApJ...779...58R} extended
the model  grid toward the high--mass domain.  They  computed five new
full evolutionary  sequences with  initial masses on  the ZAMS  in the
range  $5.5-6.7M_{\sun}$  resulting  in  WD sequences  with stellar
masses between  $0.917$  and $1.05  M_{\sun}$. 
 
The  values   of  stellar  mass  of  our complete  model grid  are
listed  in Column  1 of  Table \ref{masses}, along with  the hydrogen
(Column 2)  and helium (Column  3) content as predicted by standard
stellar  evolution, and carbon $(X_{\rm C})$ and oxygen $(X_{\rm  O})$
central abundances by mass  in Columns 4 and 5,
respectively. Additional sequences, shown in italic, were computed for
this work.  The  values of  stellar mass  of our  set  of models
covers all the  observed pulsating DA WD  stars with  a probable
carbon-oxygen core. The  maximum value of the  hydrogen envelope
(column 2), as predicted by progenitor evolution, shows a  strong
dependence  on the stellar  mass and it is determined by the limit of
H--burning.  It ranges  from   $3.2  \times  10^{-4} M_{\star}$  for
$M_{\star}  = 0.493  M_{\sun}$ to  $1.4 \times  10^{-6}  M_{\sun}$ for
$M_{\star}  =  1.050  M_{\sun}$,  with   a  value  of  $\sim  1
\times 10^{-4}M_{\star}$ for the average-mass sequence of $M_{\star}
\sim 0.60 M_{\sun}$.

\begin{table}
\caption{The main characteristics of our set of DA WD models. Sequences with the mass value in italic where computed for this work. 
The sequence with 0.493 $M_{\sun}$ comes from a full evolutionary model, while the remaining four sequences were the 
result of the interpolation process described in \citet{2013ApJ...779...58R}.}
\centering
\begin{tabular}{ccccc}
\hline\hline
$M_{\star}/M_{\sun}$ & $-\log(M_{\rm H}/M_{\star})$ & $-\log(M_{\rm He}/M_{\star})$  
& $X_{\rm C}$ & $X_{\rm O}$\\
\hline
{\it 0.493} & 3.50 & 1.08 & 0.268 & 0.720\\
0.525 & 3.62 & 1.31 & 0.278 & 0.709\\
0.548 & 3.74 & 1.38 & 0.290 & 0.697\\
{\it 0.560} & 3.70 & 1.42 & 0.296 & 0.691\\       
0.570 & 3.82 & 1.46 & 0.301 & 0.696\\
0.593 & 3.93 & 1.62 & 0.283 & 0.704\\
0.609 & 4.02 & 1.61 & 0.264 & 0.723\\
0.632 & 4.25 & 1.76 & 0.234 & 0.755\\
0.660 & 4.26 & 1.92 & 0.258 & 0.730\\
{\it 0.674} & 4.35 & 1.97 & 0.280 & 0.707\\
{\it 0.690} & 4.46 & 2.04 & 0.303 & 0.684\\
0.705 & 4.45 & 2.12 & 0.326 & 0.661\\
0.721 & 4.50 & 2.14 & 0.328 & 0.659\\
{\it 0.745} & 4.62 & 2.18 & 0.330 & 0.657\\
0.770 & 4.70 & 2.23 & 0.332 & 0.655\\
0.800 & 4.84 & 2.33 & 0.339 & 0.648\\
0.837 & 5.00 & 2.50 & 0.347 & 0.640\\
0.878 & 5.07 & 2.59 & 0.367 & 0.611\\
0.917 & 5.41 & 2.88 & 0.378 & 0.609\\
0.949 & 5.51 & 2.92 & 0.373 & 0.614\\
0.976 & 5.68 & 2.96 & 0.374 & 0.613\\
0.998 & 5.70 & 3.11 & 0.358 & 0.629\\
1.024 & 5.74 & 3.25 & 0.356 & 0.631\\
1.050 & 5.84 & 2.96 & 0.374 & 0.613\\
\hline\hline
\label{masses}
\end{tabular}
\end{table}

\begin{figure}
\begin{center}
\includegraphics[clip,width=0.47\textwidth]{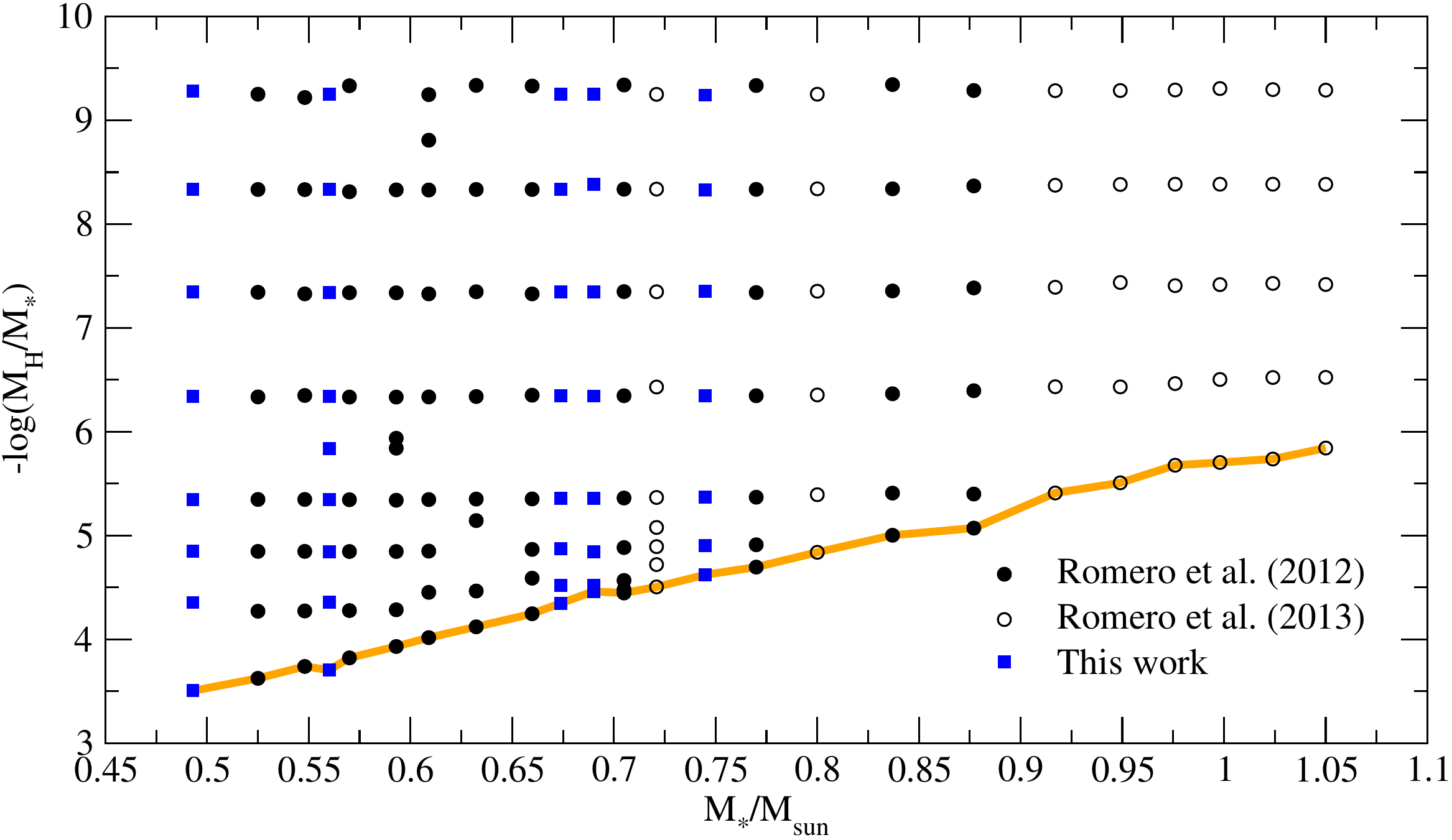}
\caption{Grid of DA WD evolutionary sequences considered in this work
  in the $M_{\star}/M_{\sun}$ vs $-\log(M_{\rm H}/M_{\star})$ plane.
  Each symbol corresponds to a sequence of models representative of
  WD stars characterized by a given stellar mass and hydrogen
  envelope mass. Filled circles correspond to the evolutionary
  sequences computed in \citet{2012MNRAS.420.1462R}, hollow circles
  correspond to sequences computed in  \citet{2013ApJ...779...58R} and
  filled squares correspond to the sequences computed in this work.
  The orange line connects the sequences with the maximum values for
  the thickness of the hydrogen envelope,  predicted by our
  evolutionary computations.}
\label{grid-masses}
\end{center}
\end{figure}

Our parameter space  is build up by varying  three quantities: stellar
mass   $(M_{\star})$,  effective   temperature  $(T_{\rm   eff})$  and
thickness  of the hydrogen  envelope $(M_{\rm  H})$. Both  the stellar
mass and the  effective temperature vary consistently  as a result of
the use of a fully evolutionary approach. On  the other  hand, we
decided to  vary the  thickness of the  hydrogen envelope in  order to
expand our parameter  space. The choice of varying  $M_{\rm H}$ is not
arbitrary, since  there are uncertainties  related to physical
processes operative during the TP-AGB phase leading  to  uncertainties
on the  amount  of hydrogen  remaining on  the  envelope of  WD  stars
\citep[see][for  a detailed              justification              of
  this
  choice]{2012MNRAS.420.1462R,2013ApJ...779...58R,2015A&A...576A...9A}. 
In  order to get  different values  of the  thickness of  the hydrogen
envelope,     we     follow     the     procedure     described     in
\citet{2012MNRAS.420.1462R,2013ApJ...779...58R}.   
For  each  sequence
with a given stellar mass and  a thick H envelope, as predicted by the
full computation of  the pre-WD evolution (Column  2 in Table
\ref{masses}),  we replaced  $^1$H with  $^4$He at  the bottom  of the
hydrogen  envelope.  This  is  done  at  high  effective  temperatures
($\lesssim  90\,000$  K), so  the  transitory  effects  caused by  the
artificial procedure are completely  washed out when the model reaches
the  ZZ  Ceti instability  strip. 
The  resulting  values of  hydrogen
content for different envelopes  are shown in Figure \ref{grid-masses}
for each mass. The orange thick line connects the values of $M_{\rm H}$ 
predicted by our stellar evolution (Column 2, Table \ref{masses}).

Other structural parameters do not change considerably  according to
standard  evolutionary computations.   For example,
\citet{2012MNRAS.420.1462R} showed  that the remaining helium content
of DA WD stars can be slightly lower (a factor of $3-4$) than that
predicted by standard stellar evolution only at the expense of an
increase in mass of the hydrogen-free core $(\sim 0.2M_{\sun})$.  The
structure of the carbon-oxygen chemical profiles is  basically fixed
by  the evolution during the  core helium burning stage and is not
expected to vary during the following single star evolution (we do not
consider possible merger episodes). The chemical structure of  the
carbon-oxygen core is affected  by the uncertainties inherent  to the
$^{12}$C$(\alpha,\gamma)^{16}$O reaction  rate. A detailed assessing
of the impact of this  reaction rate on the precise shape of the core
chemical structure and the pulsational properties is presented by
\citet{2017A&A...599A..21D}.

Summarizing,  we have available  a grid of $\sim  290$ evolutionary
sequences characterized  by a detailed  and updated input  physics, in
particular,  regarding  the internal  chemical  structure,  that is  a
crucial aspect for WD asteroseismology.

\subsection{Pulsation computations}
\label{pulsations}

In this study the adiabatic pulsation periods of nonradial $g$-modes
for  our complete set of DA WD  models were computed  using  the
adiabatic version  of the  {\tt LP-PUL} pulsation code described in
\citet{2006A&A...454..863C}. This  code is based on the  general
Newton-Raphson technique that solves the full fourth--order set of
equations and  boundary conditions governing linear,  adiabatic,
non-radial  stellar  oscillations   following  the dimensionless
formulation of  \citet{1971AcA....21..289D}. We used the so-called
``Ledoux-modified''  treatment to  assess  the  run of  the
Brunt-V\"aisal\"a  frequency  \citep[$N$; see][]{1990ApJS...72..335T},
generalized to include the  effects of having three different chemical
components varying  in abundance. This  code is coupled with  the {\tt
  LPCODE}  evolutionary  code.   

The  asymptotic  period  spacing  is computed as in
\citet{1990ApJS...72..335T}:

\begin{equation}
\Delta \Pi_{\ell}^{\rm a}= \frac{2\pi^2}{\sqrt{\ell(\ell+1)}} 
\left[\int_{r_1}^{r_2} \frac{N}{r} dr \right]^{-1}
\label{pspacing}
\end{equation}

\noindent where $N$ is the Brunt-V\"is\"al\"a frequency, and $r_1$ and
$r_2$ are the radii of  the inner and outer boundary  of the
propagation region, respectively. When a fraction of the core is
crystallized, $r_1$ coincides with the radius of the crystallization
front, which is moving outward as the star cools down, and the
fraction of crystallized mass increases.

We computed adiabatic pulsation $g$-modes with $\ell = 1$ and 2 and
periods in the range 80--2000 s. This range of periods corresponds (on
average) to $1\lesssim k \lesssim 50$ for $\ell= 1$ and $1\lesssim k
\lesssim 90$ for $\ell= 2$.

\section{Asteroseismological results}
\label{astero}

For our target stars, KIC 4552982, KIC 11911480, J113655.17+040952.6 and GD
1212, we searched for an asteroseismological representative model
that best matches the observed periods of each star.  To this end, we
seek for the theoretical model that minimizes the quality function
given by \citet{2009MNRAS.396.1709C}:

\begin{equation}
S = \frac{1}{N}\sqrt{\sum^{N}_{i=1} \frac{[\Pi_k^{th} -\Pi_i^{obs}]^2 \times A_i }{\sum^N_{i=1}Ai}}
\label{qbarbara}
\end{equation}

\noindent where $N$  is the  number  of the  observed periods  in the
star under study, $\Pi_k^{\rm th}$ and $\Pi_i^{\rm obs}$  are the
theoretical and observed periods, respectively and $A_i$ is the
amplitude of the observed mode.
The numerical uncertainties for $M_{\star}, T_{\rm eff}$, and
$\log(L_{\star}/L_{\sun})$ were computed by using the following
expression \citep{1986ApJ...305..740Z,2008MNRAS.385..430C}:

\begin{equation}
\sigma_j^2= \frac{d_j^2}{(S-S_{\rm 0})},
\end{equation}

\noindent where $S_{\rm 0}\equiv  \Phi(M_{\star}^{\rm 0}, M_{\rm
  H}^{\rm 0}, T_{\rm eff}^{\rm  0})$ is  the minimum of the quality
function $S$ which is  reached at $(M_{\star}^{\rm 0}, M_{\rm H}^{\rm
  0},T_{\rm eff}^{\rm 0} )$ corresponding to the best-fit  model, and
$S$ is the value of the quality function when we change the parameter
$j$  (in this case, $M_{\star}, M_{\rm  H}$, or $T_{\rm eff}$) by an
amount $d_j$, keeping  fixed the other parameters.  The quantity $d_j$
can  be evaluated  as  the  minimum step  in  the  grid of  the
parameter $j$.    The uncertainties in the other quantities
($L_{\star}, R_{\star}, g$, etc)  are derived from the uncertainties
in $M_{\star}$ and $T_{\rm eff}$.  These uncertainties represent the
internal errors of the fitting procedure. 


\subsection{KIC 11911480}

The DA WD star KIC 11911480 was discovered to be variable  from
ground-based observations as a part of the RATS-Kepler  survey
\citep{2014MNRAS.437..132R}. These observations revealed a  dominant
periodicity of $\sim 290$ s. The star was  observed by \emph{Kepler}
in the short-cadence mode in  quarters 12 and 16 (Q12 and Q16) and a
total of 13 periods were detected   \citep[see Table 2
  of][]{2014MNRAS.438.3086G}. Of these,  5 periods were identified as
$m= 0$ components of rotational triplets and the remainder ones as
$m= \pm 1$ components.  \citet{2014MNRAS.438.3086G} also determine the
spectroscopic values of the atmospheric parameters using spectra from
the double-armed Intermediate resolution Spectrograph (ISIS) on the
William Herschel Telescope (WHT) and the pure hydrogen atmosphere
models, with MLT/$\alpha$ = 0.8, from \citet{2010MmSAI..81..921K}. As
a result, they obtained $T_{\rm eff} = 12\, 160 \pm 250$ K and $\log g
=7.94 \pm 0.10$, after applying the 3D convection correction from
\citet{2013A&A...559A.104T}.  By employing our set of DA WD
evolutionary tracks, we derive $M_{\star}=  0.574 \pm 0.05
M_{\sun}$. \cite{2016MNRAS.457.2855G} determine the atmospheric
parameter using the same spectra but considering the atmosphere models
from \cite{2011ApJ...730..128T} with MLT/$\alpha$=0.8. The result was
$T_{\rm eff}= 11\,580 \pm 140$ K and $\log g = 7.96 \pm 0.04$, also
corrected by 3D convection. From these parameters we obtain a stellar
mass of $M_{\star}=  0.583 \pm 0.02 M_{\sun}$. The ``hot'' solution
obtained by \citet{2014MNRAS.438.3086G} is in better agreement with
the short periods observed in this star.

\begin{table}
\caption{Columns 1,2 and 3: The observed $m= 0$ periods of KIC
  11911480 to be employed as input  of our asteroseismological
  analysis, with the $\ell$ value fixed by the detection of rotational
  splitting components. Columns: 4, 5, 6 and 7: The theoretical
  periods with their corresponding harmonic degree, radial order and
  rotation coefficient for our best fit model for KIC11911480.}
\centering
\begin{tabular}{ccc|cccc}
\hline\hline
\noalign{\smallskip}
\multicolumn{3}{c}{Observations} & \multicolumn{4}{c}{Asteroseismology}\\
\hline
$\Pi_i^{\rm obs}$ [s] & $A_i$ [mma] & $\ell$ & $\Pi_i^{\rm Theo}$ & $\ell$ & $k$ & $C_{\rm k\ell}$\\
\noalign{\smallskip}
\hline
\noalign{\smallskip}
290.802 & 2.175 & 1 & 290.982 & 1 & 4 & 0.44332\\
259.253 & 0.975 & 1 & 257.923 & 1 & 3 & 0.47087 \\ 
324.316 & 0.278 & 1 & 323.634 & 1 & 5 & 0.36870\\
172.900 & 0.149 & - & 170.800 & 2 & 4 & 0.14153\\
202.569 & 0.118 & - & 204.085 & 2 & 5 & 0.12244\\
\noalign{\smallskip}
\hline\hline
\label{modos}
\end{tabular}
\end{table}

In our analysis, we employ only the five periods shown in Table
\ref{modos}, which correspond to the five $m= 0$ observed periods of
Q12 and Q16.  The quoted amplitudes are those of Q16. We assume that
the three large amplitude modes with periods 290.802 s, 259.253 s, and
324.316 are dipole modes because they are unambiguously identified
with the central components of triplets  ($\ell= 1$). 

\begin{figure}
\begin{center}
\includegraphics[clip,width=0.45\textwidth]{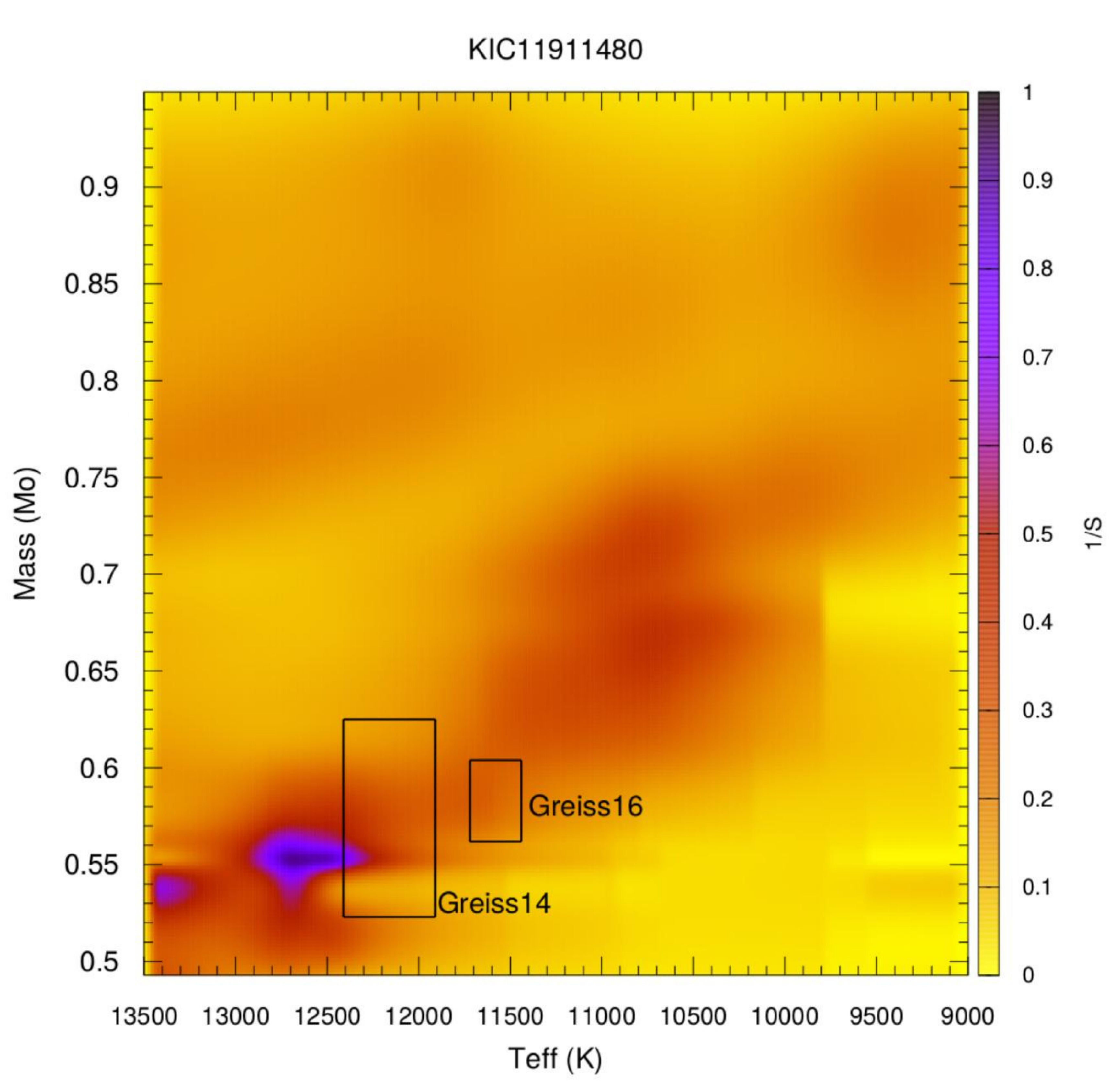}
\caption{Projection on the effective temperature vs. stellar mass
  plane of the inverse of the quality function S for KIC11911480. The
  hydrogen envelope thickness value for each stellar mass corresponds
  to the sequence with the lowest value of the quality function for
  that stellar mass. The box indicates the stellar mass and effective
  temperature values obtained from spectroscopy by
  \citet{2016MNRAS.457.2855G}.}
\label{fig1}
\end{center}
\end{figure}

Our results are shown in Figure \ref{fig1} which shows the projection
of the inverse of the quality function S on the $T_{\rm
  eff}-M_{\star}/M_{\sun}$ plane. The boxes correspond to the spectroscopic
determinations from \citet{2014MNRAS.438.3086G} and
\citet{2016MNRAS.457.2855G}. For each stellar mass, the value of the
hydrogen envelope thickness corresponds to the sequence with the lower
value of the quality function for that stellar mass. The color bar on
the right indicates the value of the inverse of the quality function
$S$. The asteroseismological solutions point to a stellar mass between
0.54 and 0.57$M_{\sun}$, with a blue edge-like effective temperature,
in better agreement with the spectroscopic determination from
\citet{2014MNRAS.438.3086G}, as can be seen from Figure \ref{fig1}.
The parameters of the model characterizing the minimum of $S$ for KIC
11911480 are listed in Table \ref{tabla-soluciones}, along with the
spectroscopic parameters. Note that the seismological effective
temperature is quite high, even higher than the classical blue edge of
the instability strip \citep{2011ApJ...743..138G}. However, the
extension of the instability strip is being redefined with some ZZ
Ceti stars characterized with high effective temperatures. For
instance, \citet{2017ApJ...841L...2H} reported the existence of the
hottest known ZZ Ceti, EPIC 211914185, with $T_{\rm eff}= 13\,590 \pm
340$ and $M_{\star} = 0.87 \pm 0.03 M_{\sun}$. Also, we can be
overestimating the effective temperature obtained from
asteroseismology.

\begin{table*}
\caption{List of parameters characterizing the best fit model obtained
  for KIC 11911480. Also, we list the spectroscopic values from
  \citet{2014MNRAS.438.3086G} and \citet{2016MNRAS.457.2855G}. The
  quoted uncertainties are the intrinsic uncertainties of the
  seismological fit. } \centering
\begin{tabular}{ccc}
\hline\hline
 \citet{2014MNRAS.438.3086G} &\citet{2016MNRAS.457.2855G} & LPCODE\\
\hline
$M_{\star} = 0.574\pm 0.05 M_{\sun}$   & $M_{\star} = 0.583\pm 0.05 M_{\sun}$ & $M_{\star} = 0.548\pm 0.01 M_{\sun}$\\
 $T_{\rm eff}= 12\, 160\pm 250$ K & $T_{\rm eff}= 11\, 580\pm 140$ K &  $T_{\rm eff} = 12\, 721\pm 228$ K \\
 $\log g = 7.94\pm 0.10$        & $\log g = 7.96\pm 0.04$     & $\log g = 7.88\pm 0.05$\\ 
		                && $\log(L/L_{\sun}) = -2.333 \pm 0.032$ \\
                             && $R/R_{\sun}=0.014 \pm 0.001$\\ 		             
                             &&  $M_{\rm H}/M_{\sun}= 2.088\times 10^{-4}$   \\
                             && $M_{\rm He}/M_{\sun}=4.19\times 10^{-2}$   \\
                             && $X_{\rm C} = 0.290, X_{\rm O} =0.697$  \\
                             && $P_{\rm rot}= 3.36 \pm 0.2$ d \\
                             && $S= 1.13$ s                \\    
\hline\hline
\label{tabla-soluciones}
\end{tabular}
\end{table*}

The list of theoretical periods corresponding to the model in Table
\ref{tabla-soluciones} is shown in Table \ref{modos}. Also listed are
the harmonic degree, the radial order and the $C_{\rm k\ell}$ rotation
coefficient. Using the frequency spacing $\Delta f$ for the three
$\ell =1$ modes from Table 2 of \citet{2014MNRAS.438.3086G} and the
rotation coefficients we estimated a rotation period of $3.36 \pm 0.2$
days. 


\subsection{J113655.1+040952.6}
\label{binary}

J1136$+$0409 (EPIC 201730811) was first  observed  by
\citet{2015MNRAS.447..691P} as part of a  search for ZZ Ceti  stars
among the  WD $+$ MS  binaries and it turn out to be the only variable
in a post common  envelope binary from the sample studied by these
authors.  This star was spectroscopically identified as a  WD $+$ dM
from its SDSS spectrum.  The surface parameters  were determined  by
\citet{2012MNRAS.419..806R} by model-atmosphere fits to the Balmer
absorption lines after subtracting an M star spectrum,  giving $T_{\rm
  eff} = 11\, 700 \pm 150$ K  and $\log g = 7.99  \pm
0.08$. Pulsations were  confirmed by a short  run  with the  ULTRACAM
instrument  mounted  on the  3.5m  New Technology  Telescope  by
\citet{2015MNRAS.447..691P}.  \citet{2015MNRAS.451.1701H} reported the
results from a  78 days  observation run  in August 2014 with  the
\emph {Kepler} spacecraft in the frame of the extended \emph {Kepler}
mission, K2 Campaign 1.   In addition, these  authors obtained high
S/N spectroscopy with SOAR to  refine the determinations of the
atmospheric parameters. They used  two independent  grids of
synthetic spectra  to fit  the Balmer lines: the  pure hydrogen
atmosphere models and  fitting procedure described by
\citet{2011ApJ...743..138G}, and the pure  hydrogen atmosphere  models
from \citet{2010MmSAI..81..921K}. Both grids employ the ML2/$\alpha=
0.8$ prescription of the mixing-length theory
\citep{2011ApJ...743..138G}.   By applying the  3D correction from
\citet{2013A&A...559A.104T}  they obtained  $T_{\rm eff}= 12\, 579 \pm
250$  K and $\log g = 7.96 \pm  0.05$  for the values obtained with
the  \citet{2011ApJ...743..138G}   fit  and $T_{\rm  eff}= 12\,  083\pm  250$ K  
and $\log  g =  8.02 \pm  0.07$   for the
\citet{2010MmSAI..81..921K} fit. From  these values, we computed the
stellar mass of J113655.17+040952.6 by employing our set of evolutionary
sequences, and  obtained $M_{\star} = 0.585  \pm 0.03 M_{\sun}$  and
$M_{\star}= 0.616  \pm 0.06  M_{\sun}$, respectively. Recently, \citet{2017arXiv170907004H} determined the atmospheric parameters using the same spectra as \citet{2015MNRAS.451.1701H} and the MLT$/\alpha$=0.8 models from \citet{2011ApJ...730..128T}, resulting in $T_{\rm eff}= 12\, 480 \pm 170$  K and $\log g = 7.956 \pm 0.0435$, similar to those obtained by using the model grid from \citet{2011ApJ...743..138G}.
As in the case of KIC 11911480, in our analysis we consider both spectroscopic
determinations from \citet{2011ApJ...743..138G} and \citet{2010MmSAI..81..921K} with the corresponding 3D correction.

\begin{table}
\caption{Columns 1,2 and 3: Observed  periods of J113655.17+040952.6 to be
  employed  as input of our asteroseismological analysis with the
  $\ell$ value fixed for three modes, according to
  \citet{2015MNRAS.451.1701H}. Columns 4, 5, 6 and 7: The theoretical
  periods with their corresponding harmonic degree, radial order and
  rotation coefficient for our best fit model for J113655.17+040952.6.}
\centering
\begin{tabular}{ccc|cccc}
\hline\hline
\multicolumn{3}{c}{Observation} & \multicolumn{4}{c}{Asteroseismology}\\
 $\Pi_i^{\rm obs}$ & $A_i$ (ppt) & $\ell$ & $\Pi_i^{\rm Theo}$ & $\ell$ & $k$ & $C_{\rm k\ell}$\\
\hline
279.443 & 2.272 & 1 & 277.865 & 1 & 3 & 0.44222\\
181.283 & 1.841 & - & 185.187 & 1 & 2 & 0.37396\\
162.231 & 1.213 & 1 & 161.071 & 1 & 1 & 0.48732\\
344.277 & 0.775 & 1 & 344.218 & 1 & 5 & 0.47552\\
201.782 & 0.519 & - & 195.923 & 2 & 4 & 0.14507\\
\hline
\hline
\label{J1136-obs}
\end{tabular}
\end{table}

From the  analysis of the  light curve, \citet{2015MNRAS.451.1701H}
found 12 pulsation frequencies, 8 of them being components  of
rotational triplets ($\ell= 1$). Only 7 frequencies were
identified with $m=0$ components. Further analysis of the light curve
revealed that the two modes with the lower amplitudes detected were
not actually real modes but nonlinear combination frequencies. We
consider 5 periods for our asteroseismic study, which are listed in
Table \ref{J1136-obs}.  According to \citet{2015MNRAS.451.1701H},  the
modes with periods 279.443  s, 162.231  s and 344.277 s are the
central $m= 0$   components of rotational  triplets. In particular,
the 344.277 s period is not detected  but it corresponds to the mean value of the frequencies of 2848.17 and 2761.10 $\mu$Hz, identified as the prograde and retrograde components, respectively. We   assume that the harmonic degree of the
periods identified as $m= 0$ components of triplets
\citep{2015MNRAS.451.1701H} is $\ell= 1$.

\begin{table*}
\caption{List of parameters characterizing the best fit model obtained
  for J113655.17+040952.6 along with the spectroscopic determinations from
  \citet{2015MNRAS.451.1701H} using the atmosphere models from
  \citet{2011ApJ...743..138G} (G2011) and  \citet{2010MmSAI..81..921K}
  (K2010). The quoted uncertainties are the intrinsic uncertainties of
  the seismological fit. } \centering
\begin{tabular}{ccc}
\hline\hline
\multicolumn{2}{c}{\citet{2015MNRAS.451.1701H}} & LPCODE\\
G2011 & K2010 & \\
\hline
$M_{\star} = 0.585\pm 0.03 M_{\sun}$ & $M_{\star} = 0.616\pm 0.06 M_{\sun}$ & $M_{\star} = 0.570 \pm 0.01 M_{\sun}$\\
$T_{\rm eff}= 12\, 579\pm 250$ K & $T_{\rm eff}= 12\, 083\pm 250$ K &  $T_{\rm eff} = 12\, 060 \pm 300$ K \\
$\log g = 7.96\pm 0.05$  &    $\log g = 8.02\pm 0.07$  & $\log g =7.95\pm 0.07$\\
                            & & $\log(L/L_{\sun})= -2.414\pm 0.045$\\
                            & & $R/R_{\sun} = 0.0132\pm 0.002$\\
                            & &  $M_{\rm H}/M_{\sun}= 1.774\times 10^{-5}$   \\
                            & & $M_{\rm He}/M_{\sun}=3.50\times 10^{-2}$   \\
                            & & $X_{\rm C}=0.301 , X_{\rm O} =0.696$  \\
                            & & $P_{\rm rot} = 2.6 \pm 1$ hr \\
                            & & $S= 2.83$ s                \\    
\hline\hline
\label{sol-J1136}
\end{tabular}
\end{table*}

\begin{figure}
\begin{center}
\includegraphics[clip,width=0.45\textwidth]{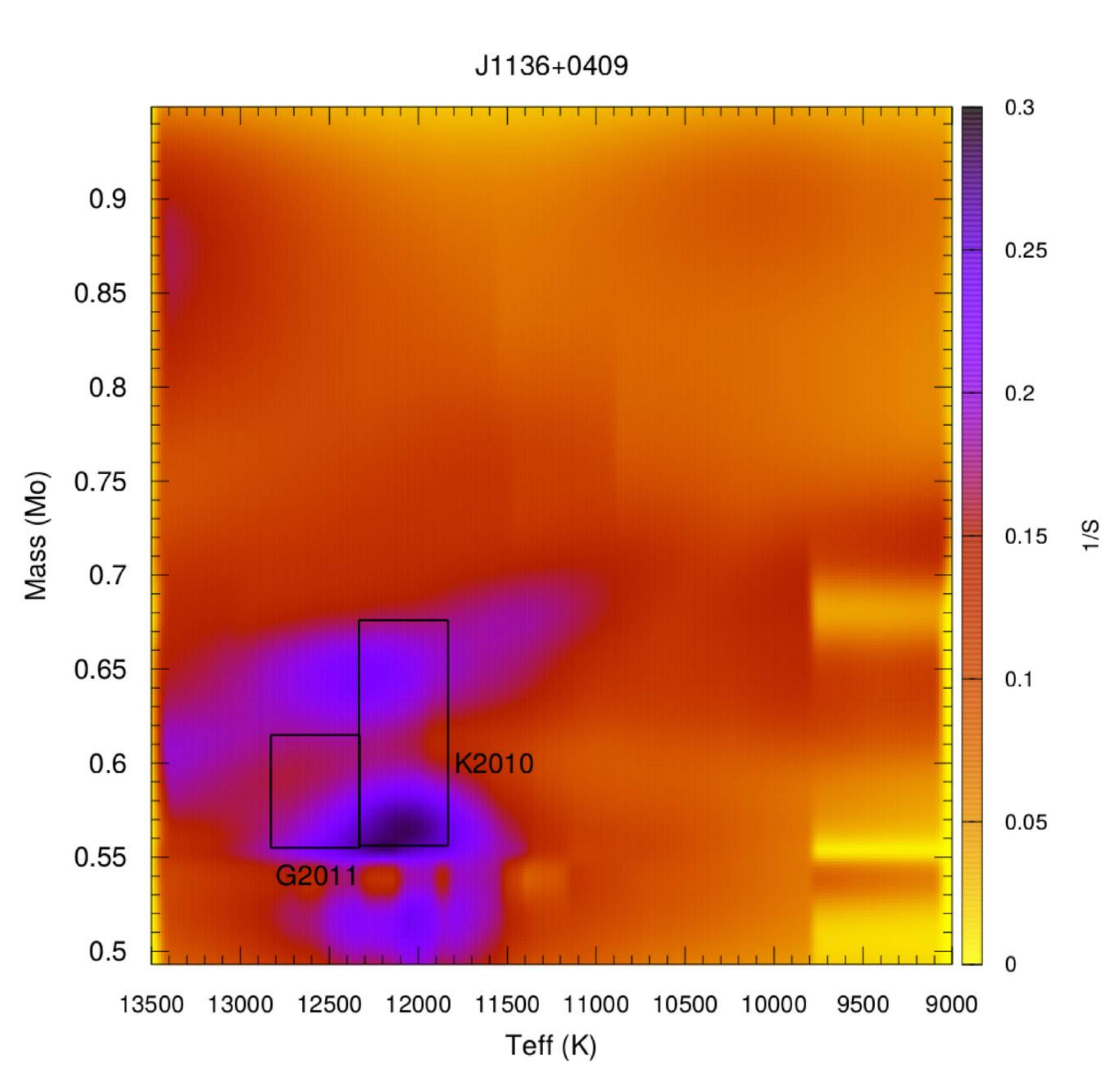}
\caption{Projection on the effective temperature vs. stellar mass
  plane of the inverse of the quality function $S$ for J113655.17+040952.6. The
  box indicates the spectroscopic determinations from
  \citet{2015MNRAS.451.1701H}.}
\label{mapa-J1136}
\end{center}
\end{figure}

The results for our asteroseismological fits are shown in figure
\ref{mapa-J1136}, which shows the  projection  of the inverse of the
quality function $S$ on the $T_{\rm eff}-M_{\star}/M_{\sun}$ plane. The
hydrogen envelope thickness value for each stellar mass corresponds to
the sequence with the lowest value of the quality function. We show
the spectroscopic values from \citet{2015MNRAS.451.1701H} with
boxes. As can be seen from this figure, we have a family of minimum
around $\sim 0.57 M_{\sun}$ and $12\, 000$ K. The structural
parameters characterizing the best fit model are listed in Table
\ref{sol-J1136} while the list of theoretical periods are listed in
the last four columns of Table \ref{J1136-obs}. Note that, in addition
to the three modes  for which we fixed the harmonic degree to be $\ell
=1$  (279.443 s, 162.231 s, and 344.407 s),  the mode  with period
181.283 s,  showing  the second largest amplitude, is also fitted by a
dipole theoretical mode.   Our seismological  stellar mass is somewhat
lower than the values shown in Table \ref{J1136-obs}, but  still
compatible with  the spectroscopic  determinations. The
effective temperature is a blue edge-like value closer to the
determinations using  \citet{2010MmSAI..81..921K} atmosphere models.
In  addition,  we obtain  a  hydrogen envelope $\sim  20\%$ thicker
than the seismological results presented in
\citet{2015MNRAS.451.1701H}. Since the central oxygen  composition is
not a free  parameter in our  grid, the oxygen   abundance at the core
of the WD model is fixed  by the previous  evolution, and has a value
of $X_{\rm O} = 0.696$, much lower than the value found by
\citet{2015MNRAS.451.1701H} of $X_{\rm O} = 0.99$. Note that even
taking into account the uncertainties in the
$^{12}$C$(\alpha,\gamma)$O$^{16}$ reaction rate given in
\citet{2002ApJ...567..643K} the abundance of oxygen can only be as
large as $X_{\rm O}=0.738$ \citep{2017A&A...599A..21D}. Results from 
\citet{2017arXiv170903144D} are also consistent with a $\sim$10\% 
uncertainty in the oxygen central abundance. Finally, we
computed the rotation coefficients $C_{k\ell}$ (last column in Table
\ref{J1136-obs}) and used the identified triplets to derived a mean
rotation period of $2.6 \pm 0.1$ hr. 
 
\subsection{KIC 4552982}
\label{KeatonBell}

KIC 4552982,  also known as SDSS J191643.83+393849.7, was identified
in the  \emph {Kepler}  Mission field through   ground-based   time
series   photometry  by  \citet{2011ApJ...741L..16H}. These authors
detected seven frequencies of  low-amplitude luminosity variations
with periods  between $\sim 800$ s  and $\sim 1450$ s.  The stellar
mass  and effective temperature  determinations are $T_{\rm eff}= 11\, 129 \pm 115$ K and $\log g = 8.34\pm 0.06$ that corresponds to  $M_{\star}= 0.82 \pm 0.04  M_{\sun}$. By applying the 3D convection correction \citet{2015ApJ...809...14B} obtained  $T_{\rm eff}= 10\, 860 \pm 120$ K and $\log g = 8.16\pm 0.06$ that corresponds to  $M_{\star}= 0.693 \pm 0.047 M_{\sun}$. Similar results were reported by \citet{2017arXiv170907004H} using the same spectra and the model grid from \citet{2011ApJ...730..128T},  $T_{\rm eff}= 10\, 950 \pm 160$ K, $\log g = 8.113\pm 0.053$ and $M_{\star}= 0.665 \pm 0.030 M_{\sun}$.

\citet{2015ApJ...809...14B} presented photometric data for
KIC 4552982 spanning  more than 1.5 years obtained with \emph
{Kepler}, making it the  longest pseudo-continuous light  curve ever
recorded for a ZZ Ceti star. They identify 20 periods from $\sim 360$
s to $\sim 1500$  s (see Table  \ref{KIC45-obs}). From the list, it is
apparent that the three modes  around $\sim 361$ s   are very close,
and probably they are  part of a $\ell= 1$ rotation  multiplet
\citep{2015ApJ...809...14B}.  Therefore, we can consider  the
observed period of $361.58$ s as the $m= 0$ component of the triplet
and assume that this period is associated to a dipole ($\ell= 1$)
mode.  \citet{2015ApJ...809...14B} have searched for a possible period
spacing in their list of periods. They found two sequences
with evenly  space periods, being the period separations of $41.9  \pm 0.2$  s 
and $20.97  \pm 0.02$, identified as possible $\ell =1$ and
$\ell=2$  sequences, respectively. By using the strong  dependence of
the asymptotic period spacing with the stellar mass, we can estimate the stellar
mass of  KIC 4552982 as $M_{\star} \sim 0.8M_{\sun}$ and thick
hydrogen envelope. 

\begin{table}
\caption{Observed periods of KIC 4552982 according to
  \citet{2015ApJ...809...14B}. The amplitudes correspond to the square
  root of the Lorentzian height listed in Table 2 of
  \citet{2015ApJ...809...14B}. Column 3 shows the theoretical periods correspondign to the Best fit model (BFM) (see. Table \ref{sol-KIC455} or first row in Table \ref{sol-sip}) with the corresponding harmonic degree and radial order ($\ell, k$). Column 4 list the theoretical periods, and ($\ell$,$k$), for the second best fit model (see second row of Table \ref{sol-sip}). }
\centering
\begin{tabular}{cc|c|c}
\hline\hline
$\Pi_i^{\rm obs}$ & $A_i$ (mma) & $\Pi_i^{\rm Theo}$ (BFM) & $\Pi_i^{\rm Theo}$ \\
\hline
 360.53 &  $\cdots$ & $\cdots$ & $\cdots$\\
 361.58 & $\cdots$ & 361.20 (1,5) & 361.25 (1,6) \\   
 362.64 & 0.161 & $\cdots$ & $\cdots$\\
 788.24 & 0.054 & 788.57 (1,14) & 788.35 (1,7)\\  
 828.29 & 0.142 & 829.27(1,15)  & 831.17 (1,18)\\  
 866.11 & 0.163 & 870.34 (1,16) & 873.94 (1,19)\\  
 907.59 & 0.137 & 907.91 (1,17) & 917.99 (1,20)\\  
 950.45 & 0.157 & 944.62 (1,18) & 949.16 (1,21)\\  
 982.23 & 0.090 & 984.00 (2,33) & 982.14 (1,22)\\  
1014.24 & 0.081 & 1018.11 (2,34) & 1021.97 (2,40) \\  
1053.68 & 0.056 & 1048.47 (2,35) & 1049.40 (2,41)\\  
1100.87 & 0.048 & 1098,72 (2,37) & 1095.46 (2,43)\\  
1158.20 & 0.074 & 1155.79 (2,39) & 1154.85 (1,26)\\  
1200.18 & 0.042 & 1201.51 (1,23) & 1200.26 (2,51)\\  
1244.73 & 0.048 & 1245.58 (1,24) & 1245.22 (2,49)\\     
1289.21 & 0.115 & 1290.06 (1,25) & 1292.77 (1,29)\\  
1301.73 & 0.084 & 1299.40 (2,44) & 1295.67 (2,51) \\  
1333.18 & 0.071 & 1333.14 (2,45) & 1340.16 (2,53)\\  
1362.95 & 0.075 & 1358.30 (2,46) & 1362.91 (1,31)\\  
1498.32 & 0.079 & 1502.55 (2,51) & 1496.03 (2,59)\\  
\hline\hline
\label{KIC45-obs}
\end{tabular}
\end{table}

We start our analysis of KIC 4552982 by  carrying out an
asteroseismological  period fit  employing  the 18  modes identified
as $m= 0$. In addition to assure that the mode with $\sim 361.6$ s is
the $m= 0$ component of a triplet,  \citet{2015ApJ...809...14B} also
identify  the modes with  period between 788 and 950 s as $\ell= 1$
modes.  These modes are separated by a nearly constant period  spacing
and have  similar  amplitudes \citep[see
  Fig. 10][]{2015ApJ...809...14B},  except  for the  mode with  788.24
s. Then we consider all five periods as dipole modes and fix the
harmonic degree to $\ell=1$.  We allow the remainder periods to be
associated to either $\ell= 1$ or $\ell= 2$ modes, without restriction
at  the outset. 

\begin{figure}
\begin{center}
\includegraphics[clip,width=0.5\textwidth]{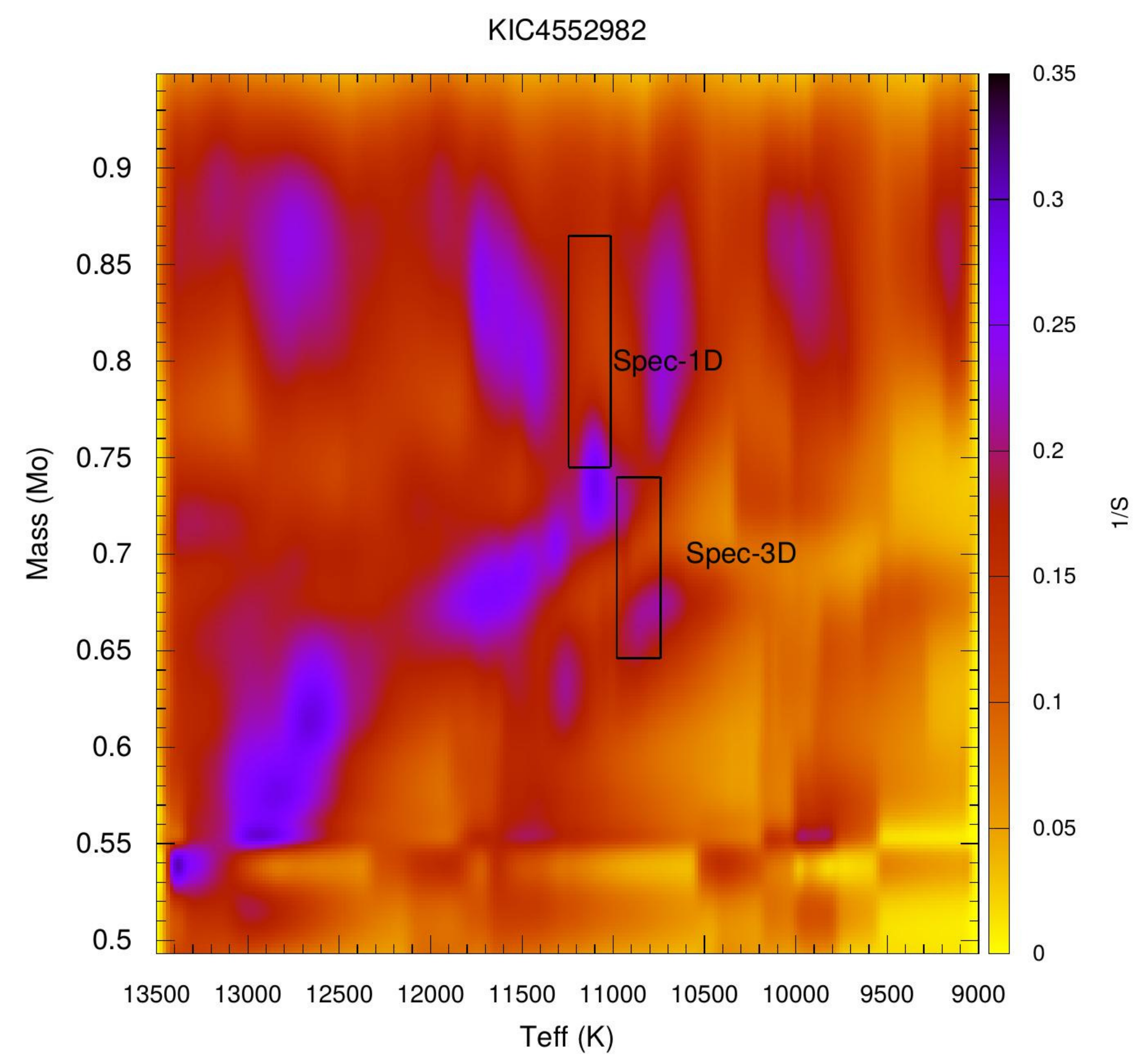}
\caption{Projection on the effective temperature vs. stellar mass of
  $1/S$ for KIC 4552982.  We fixed the harmonic degree for the six
  modes with the shortest periods ($\ell=1$). Spectroscopic
  determinations with and without the 3D convection correction are
  also depicted as boxes.}
\label{solutions-KIC}
\end{center}
\end{figure}

In Fig. \ref{solutions-KIC} we show the projection on the
$T_{\rm eff}-M_{\star}$ plane of $1/S$ corresponding to the seismological fit of
KIC 4552982.The hydrogen envelope value corresponds to the sequence
with the lowest value of the quality function for that stellar mass.
We include in the figure the spectroscopic determinations of the
effective temperature and stellar mass for KIC 4552982 with  (Spec-3D)
and  without (Spec-1D) correction from  \citet{2013A&A...559A.104T}
with the associated uncertainties as a  box. From this figure two
families of solutions can be identified: A "hot" family with $T_{\rm
  eff} > 12\, 000$K and stellar mass between 0.55 and 0.65$M_{\sun}$
and "cool" family with $T_{\rm eff}\sim 11\, 000$K and stellar mass
$\sim 0.72 M_{\sun}$. This star has a rich period spectra, with 18
pulsation periods showing similar amplitudes. Then, with no
amplitude--dominant mode, the period spacing would have a strong
influence on the quality function and thus in the seismological fit.
Note that the asymptotic period spacing increases with decreasing mass
and effective temperature, then the strip in figure
\ref{solutions-KIC} formed by the two families correspond to a
"constant period spacing" strip. We disregard the "hot" family of
solutions based on the properties of the observed period spectrum,
with many long excited periods with high radial order, which is
compatible with a cool ZZ Ceti star. In addition, a high $T_{\rm eff}$
is in great disagreement with the spectroscopic determinations, as can
be seen from Fig. \ref{solutions-KIC}.

The parameters of our best fit model for KIC 4552982 are listed in
Table \ref{sol-KIC455}, along with the spectroscopic determinations
with and without the 3D convection correction. This solution is in
better agreement with the spectroscopic determinations without the
3D-corrections, as can be see from figure \ref{solutions-KIC}. Using
the data from the frequency separation for rotational splitting of
$\sim 10 \mu Hz$ and the corresponding rotation coefficient $C_{k\ell}
=0.48612$ we obtain a rotation period of $\sim 15\pm 1$ h.  The list
of theoretical periods and their values of $\ell$ and $k$
corresponding to this model are listed in the first row of Table
\ref{sol-sip}. Also listed are the asymptotic period spacing for
dipole and quadrupole modes. 
  
The model with the minimum value of the quality function within the
box corresponding to spectroscopic determinations with 3D-corrections
(Spec-3D) shows an stellar mass of $0.721 M_{\sun}$ and an effective
temperature of $10\, 875$ K. However the period-to-period fit is not
as good, with a value of the quality function of 4.87 s. The
theoretical periods for this model are listed in the second row of
Table \ref{sol-sip}.

\begin{table*}
\caption{List of parameters characterizing the best fit model obtained
  for KIC 4552982 along with the spectroscopic determinations from
  \citet{2015ApJ...809...14B} and \citet{2011ApJ...741L..16H}. The
  quoted uncertainties are the intrinsic uncertainties of the
  seismological fit. } \centering
\begin{tabular}{ccc}
\hline\hline
\citet{2011ApJ...741L..16H} & \citet{2015ApJ...809...14B}  & LPCODE\\
\hline
$M_{\star} = 0.805\pm 0.06 M_{\sun}$ &  $M_{\star} = 0.693\pm 0.047 M_{\sun}$  & $M_{\star} = 0.745 \pm 0.007 M_{\sun}$ \\
$T_{\rm eff}= 11\, 129\pm 115$ K & $T_{\rm eff}= 10\,860 \pm 120$ K & $T_{\rm eff} = 11\, 110 \pm 69$ K \\
 $\log g= 8.34\pm 0.06$ & $\log g = 8.16\pm 0.06$   & $\log g = 8.26 \pm 0.05$ \\
                             & & $\log(L/L_{\sun})=-2.815\pm 0.011$ \\
	                     & & $R/R_{\sun} = 0.0105\pm 0.0002$  \\             
                             & & $M_{\rm H}/M_{\sun}= 4.70\times 10^{-9}$  \\
                             & & $M_{\rm He}/M_{\sun}=6.61\times 10^{-3}$   \\
                             & & $X_{\rm C} = 0.330 , X_{\rm O} =0.657$\\
                             & & $P_{\rm rot}=15\pm 1$ hr\\
                             & & $S=3.45$ s  \\    
\hline\hline
\label{sol-KIC455}
\end{tabular}
\end{table*}

\begin{table} 
\caption{ Seismological solution for KIC 4552982 considering the 18
  modes identified as $m=0$ components. The harmonic degree for the
  modes with periods between 361.58 s and 950 s is fixed to be $\ell
  =1$ at the outset, in agreement with the identification and the
  possible period spacing proposed by \citet{2015ApJ...809...14B}.}
\begin{tabular}{cccccc}
\hline\hline
$M_{\star}/M_{\sun}$ & $M_{\rm H}/M_{\star}$ & $T_{\rm eff}$ [K] & $\Delta \Pi_{\ell= 1}$ &  
$\Delta \Pi_{\ell= 2}$ & $S$ (s)  \\
\hline\hline
0.745 & $4.70\times 10^{-9}$ & $11\, 110$ & 50.50 & 29.16 & 3.45   \\
0.721 & $3.13\times 10^{-5}$ & $10\, 875$ & 43.48 & 25.10 & 5.05   \\
\hline\hline
\label{sol-sip}
\end{tabular}
\end{table}

\begin{figure}
\begin{center}
\includegraphics[clip,width=0.45\textwidth]{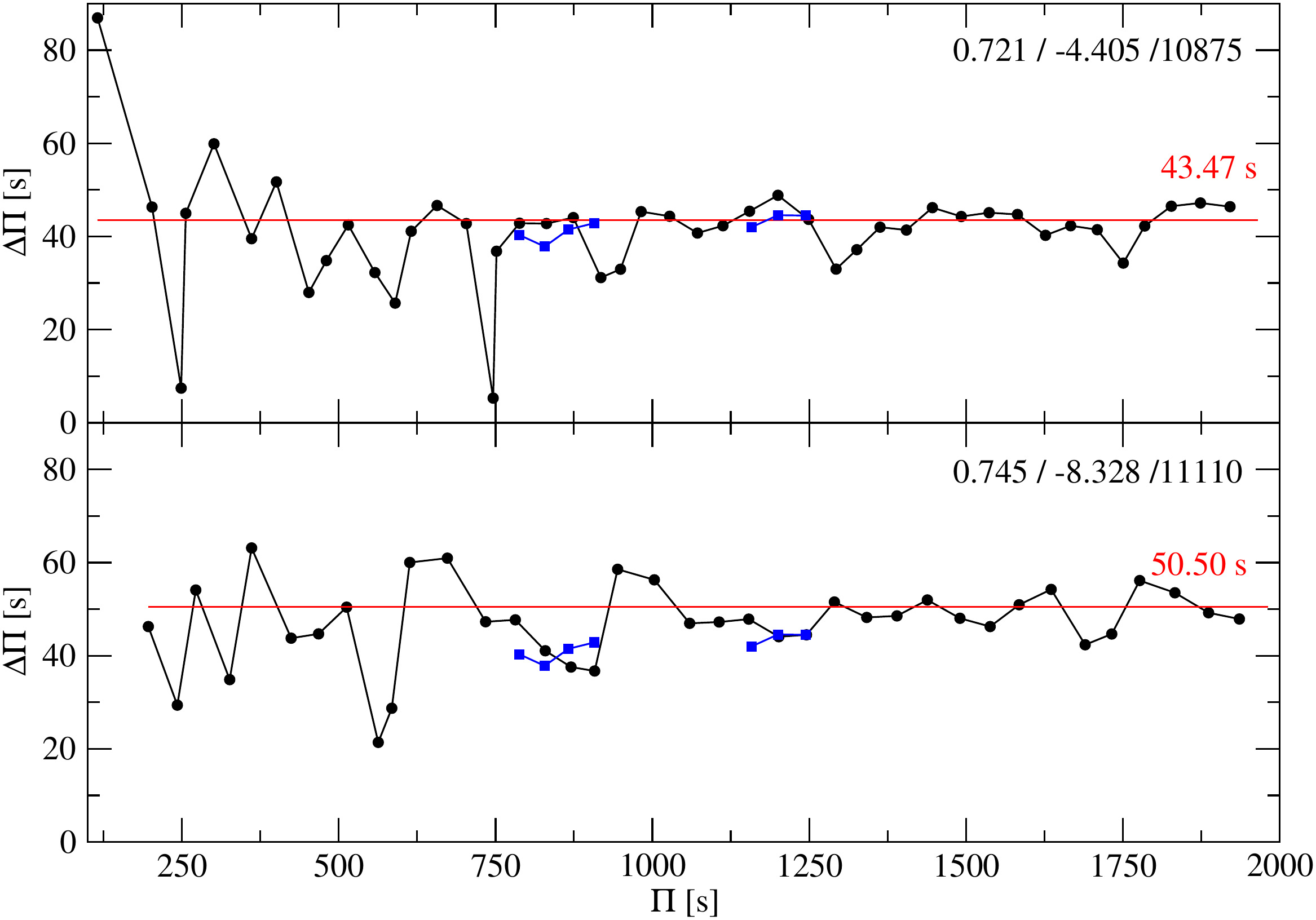}
\caption{The forward period spacing 
in terms of the periods for the theoretical models (black circles) 
listed in Table \ref{sol-sip}. In each panel we specify 
the stellar mass, the hydrogen mass  
[$\log(M_{\rm H}/M_{\star})$] and the effective temperature in K. 
The asymptotic period spacing is depicted as a red horizontal line. 
Blue squared connected with thin lines represent the 
forward period spacing of the modes identified as $\ell=1$ 
modes by  \citet{2015ApJ...809...14B}, assuming that 
these modes have consecutive radial orders.}
\label{DP-KIC}
\end{center}
\end{figure}

If we assume that the mean period  spacing of 41.9 s derived   by
\citet{2015ApJ...809...14B} for KIC 4552982 is associated to the
asymptotic period  spacing for dipole  modes,  then only the
asteroseismological  solution of $0.721 M_{\sun}$ is compatible  with
this  star.  This is illustrated  in the upper panel of
Fig. \ref{DP-KIC}, in which we depict the dipole asymptotic period
spacing (red line) for the $0.721 M_{\sun}$ model, along with the
observed forward period  spacing  ($\equiv \Pi_{k+1}-\Pi_k$) of KIC
4552982 (blue squares connected with thin lines) in terms of the
pulsation periods. In addition, the $\ell= 1$ theoretical forward
period-spacing values  are displayed with black circles. The lower
panel shows the situation for the best fit model with $M_{\star}=
0.745 M_{\sun}$. It is  apparent that for this model, the asymptotic
period  spacing is too long as to be compatible with the observed mean
period spacing of 41.9 s of KIC 4552982. However, in these cases  the
forward period spacing values of the model are  in very good agreement
with the period spacing values observed in  the star.  In summary, the
two selected models can be considered as compatible with  KIC 4552982
concerning either the mean  period spacing of 41.9 s, or the
individual forward period  spacing values exhibited by the
star. However, from the period--to--period fit the  best fit model
corresponds to that with stellar mass of $0.745M_{\sun}$ (first row in
Table \ref{sol-sip}). 

\subsection{GD 1212}

GD    1212   was    reported   to    be    a   ZZ    Ceti   star    by
\citet{2006AJ....132..831G}, showing a  dominant period at $\sim 1161$
s.  Spectroscopic  values of effective temperature and gravity from
\citet{2011ApJ...743..138G}  are $T_{\rm  eff}=11270\pm 165$  K  and
$\log  g  = 8.18  \pm 0.05$, using  their ML2/$\alpha =0.8$ atmosphere
models.  By applying the  3D corrections of
\citet{2013A&A...559A.104T} we obtain $T_{\rm eff} =  10\, 970 \pm 170$ K 
and $\log g = 8.03 \pm 0.05$. \citet{2017arXiv170907004H} determine the atmospheric parameters of GD 1212 using SOAR spectra and obtained $T_{\rm  eff}=10\, 980\pm 140$  K and
$\log  g  = 7.995  \pm 0.040$, by applying the atmosphere model grid from \citet{2011ApJ...730..128T}. The  ML2/$\alpha =0.8$ model
atmosphere   fits   to   the photometry of GD  1212 lead to a somewhat
lower effective temperature and a higher gravity, $T_{\rm eff} =  10\,
940 \pm 320$ K and $\log g = 8.25 \pm 0.03$
\citep{2012ApJS..199...29G}. By employing our set of DA WD
evolutionary tracks, we derive the stellar mass of GD 1212 from its
observed  surface  parameters,   being  $M_{\star}=  0.619  \pm  0.027 M_{\sun}$, 
$M_{\star} = 0.600 \pm 0.021 M_{\sun}$ and $M_{\star} = 0.747 \pm 0.023 M_{\sun}$, corresponding to
the two 3D corrected spectroscopic and photometric  determinations of
$T_{\rm eff}$ and $\log g$, respectively. From a  total of 254.5 hr of
observations with the \emph{Kepler}  spacecraft,
\citet{2014ApJ...789...85H} reported the detection of 19 pulsation
modes with periods  between 828.2 and  1220.8 s (see first column of 
Table  \ref{GD1212-obs}). Both the discovery periods
and  those observed with the \emph {Kepler} spacecraft are consistent
with a  red edge ZZ Ceti pulsator, with effective temperatures $\sim 11\, 000$ K.  
\citet{2017arXiv170907004H} reanalyzed the data using only the final 9 days of 
the K2 engineering data. After concluding that the star rotates with a period of $\sim 6.9$ days, they found five modes corresponding to $m=0$ components of multiples, along with two modes with no identified harmonic degree. These period values for the seven modes are listed in columns 3 and 4 of Table \ref{GD1212-obs}. 

\begin{table}
\caption{List of periods for GD 1212 corresponding to \citet{2014ApJ...789...85H} (column 1) and \citet{2017arXiv170907004H} (columns 2 and 3) }
\centering
\begin{tabular}{c|ccc}
\hline\hline
\citet{2014ApJ...789...85H} & & This work & \\
$\Pi_i^{\rm obs}$ & $\Pi_i^{\rm obs}$ & HWHM & $\ell$ \\
\hline
 $\cdots$ & 369.85 & 0.348 & ?\\
 $828.19 \pm 1.79$ & 826.26 & 0.593 & 2 \\
 $842.96 \pm 1.02$ & 842.90 & 0.456 & 1\\
 $849.13 \pm 0.76$ & $\cdots$ & $\cdots$ & -\\ 
 $857.51 \pm 0.86$ & $\cdots$ & $\cdots$ & -\\
 $871.06 \pm 2.13$ & $\cdots$ & $\cdots$ & -\\
 $956.87 \pm 4.91$ & 958.39 & 0.870 & ?\\
 $987.00 \pm 3.73$ & $\cdots$ & $\cdots$ & -\\
$1008.07 \pm 1.20$ & $\cdots$ & $\cdots$ & -\\
$1025.31 \pm 2.26$ & $\cdots$ & $\cdots$ & -\\
$1048.19 \pm 4.01$ & $\cdots$ & $\cdots$ & -\\
$1063.08 \pm 4.13$ & 1063.1 & 0.970 & 2\\
$1086.12 \pm 3.27$ & 1085.86 & 0.558 & 2\\
$1098.36 \pm 1.65$ & $\cdots$ & $\cdots$ & -\\
$1125.37 \pm 3.01$ & $\cdots$ & $\cdots$ & -\\
$1147.12 \pm 3.19$ & $\cdots$ & $\cdots$ & -\\
$1166.67 \pm 4.81$ & $\cdots$ & $\cdots$ & -\\
$1180.40 \pm 4.02$ & $\cdots$ & $\cdots$ & -\\
$1190.53 \pm 2.28$ & 1190.5 & 0.789 & 1 \\
$1220.75 \pm 7.15$ & $\cdots$ & $\cdots$ & -\\
\hline\hline
\end{tabular}
\label{GD1212-obs}
\end{table}

\begin{figure}
\begin{center}
\includegraphics[clip,width=0.47\textwidth]{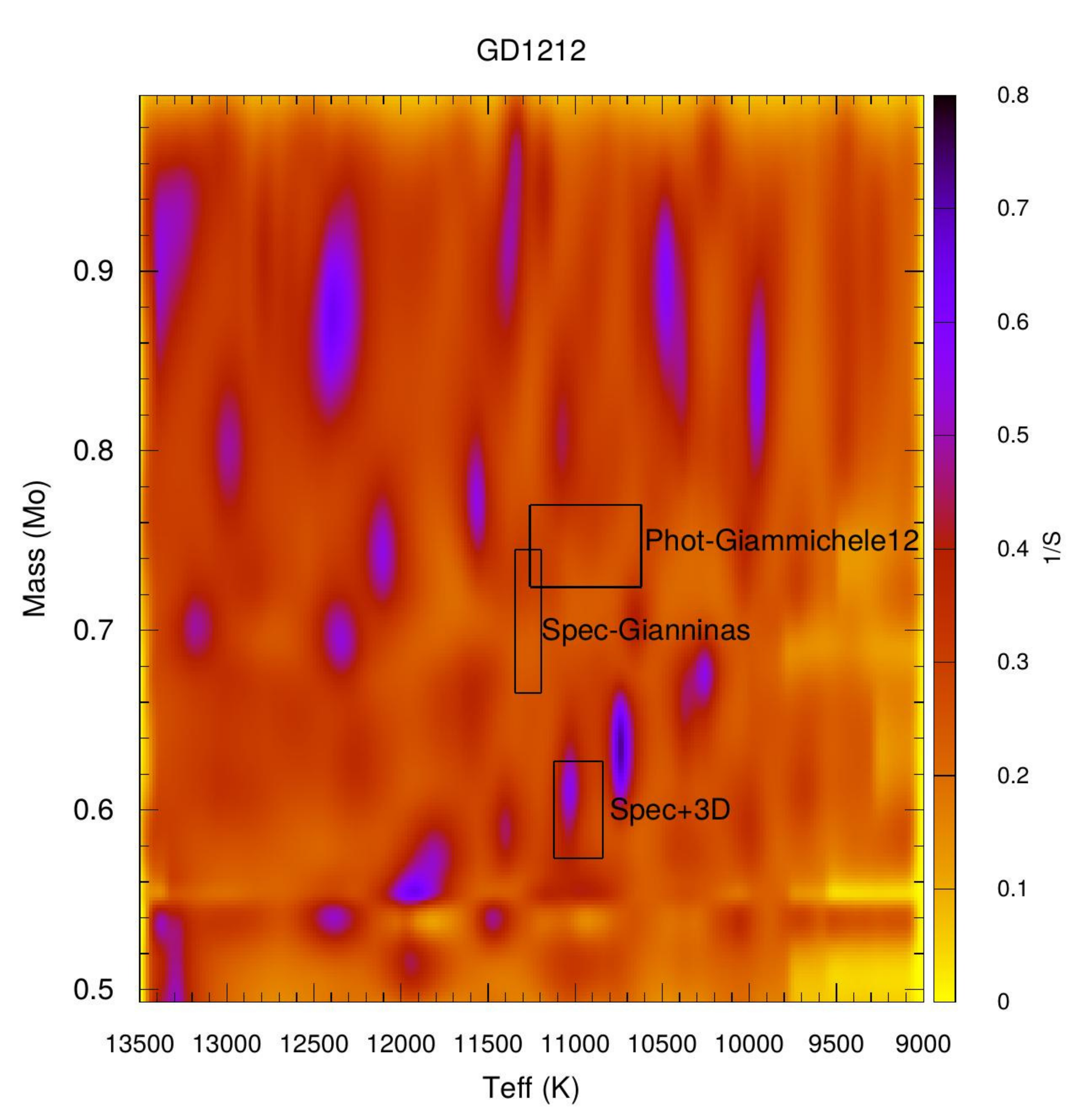}
\caption{Projection on the effective temperature vs. stellar mass
  plane of the inverse of the quality function $S$ for GD 1212. Open
  rectangles indicate the values obtained from spectroscopy
  \citet{2011ApJ...743..138G},  with 3D convection correction from
  \citet{2013A&A...559A.104T} \citep{2014ApJ...789...85H} and from
  photometry \citet{2012ApJS..199...29G}.}
\label{libre-fig}
\end{center}
\end{figure}

In this work we use the list of periods shown in the column 3 of Table
\ref{GD1212-obs} \citep{2017arXiv170907004H} to perform our asteroseismological study. 
Two modes are identified as dipole modes. Then we fix the harmonic degree to be $\ell=1$ for these modes
(see Table \ref{GD1212-obs}), and allow the remaining modes to be
associated to dipoles or quadrupoles.  To  find the  best fit models
we looked  for those models associated with minima in the quality
function $S$, to ensure that the theoretical periods are the closest
match to the observed values.  The results from our fit are shown in
Figure \ref{libre-fig}. The spectroscopic values from
\citet{2011ApJ...743..138G},  with 3D convection correction from
\citet{2013A&A...559A.104T} and from photometry
\citep{2012ApJS..199...29G} are depicted with black boxes. 
From this figure, a well defined family of solutions can be seen around $M_{\star} =0.63M_{\sun}$ and $T_{\rm eff} = 10\, 750$ K. The structure parameter
characterizing the best fit model for GD 1212 are listed in Table
\ref{GD1212-model}. The theoretical periods and the corresponding
harmonic degree and radial order are listed in Table
\ref{GD1212-teo}. Note that, appart from the two modes for which we fixed the harmonic degree to be $\ell=1$, the modes identified by \citet{2017arXiv170907004H} as $\ell=2$ modes, are also quadrupole modes in our best fit model, as the two modes with no defined harmonic degree.

\begin{table}
\caption{List of parameters characterizing the best fit model obtained
  for GD 1212 along with the spectroscopic determinations with and
  without 3D convection correction, and photometry. The quoted
  uncertainties are the intrinsic uncertainties of the seismological
  fit. } \centering
\begin{tabular}{cc}
\hline\hline
\citet{2014ApJ...789...85H} & LPCODE\\
\hline
 $M_{\star} = 0.600\pm 0.027 M_{\sun}$ &  $M_{\star} = 0.632 \pm  M_{\sun}$\\
 $T_{\rm eff}= 10\,980 \pm 140$ K &  $T_{\rm eff} = 10\, 737 \pm 70$ K \\
$\log g= 8.03\pm 0.05$       & $\log g = 8.05\pm 0.04$\\ 
                             & $\log (L/L_{\sun})= -2.737\pm 0.008$\\
                             & $R/R_{\sun} = 0.0123 \pm 0.0003$\\
                             &  $M_{\rm H}/M_{\sun}= 7.582\times 10^{-5}$   \\
                             &  $M_{\rm He}/M_{\sun}=1.74\times 10^{-2}$   \\
                             &  $X_{\rm C}= 0.234 , X_{\rm O}=0.755$  \\
                             &  $S=1.32$ s                \\    
\hline\hline
\label{GD1212-model}
\end{tabular}
\end{table}

\begin{table}
\caption{The theoretical periods with their corresponding harmonic degree and radial order for our best fit model for GD 1212.}
\centering
\begin{tabular}{ccc}
\hline\hline
$\Pi_i^{\rm Theo}$ & $\ell$ & $k$ \\
\hline
369.342 & 2 & 12 \\
826.191 & 2 & 30 \\
841.005 & 1 & 17 \\
956.400 & 2 & 35 \\
1064.42 & 2 & 39 \\
1086.32 & 2 & 40 \\
1191.45 & 1 & 25 \\
\hline
\hline
\label{GD1212-teo}
\end{tabular}
\end{table}


We also performed a seismological analysis based on the periods
reported by \citet{2014ApJ...789...85H}. Using the period spacing for
$\ell=1$  modes  of  $\Delta\Pi  =   41.5  \pm  2.5$  s determined by
\citet{2014ApJ...789...85H} and the spectroscopic effective
temperature we estimated the stellar mass by comparing this value to
the theoretical asymptotic  period spacing corresponding to canonical
sequences, listed in Table \ref{masses}.  As  a  result,  we obtained
$M_{\star}  = (0.770\pm 0.067)M_{\sun}$.  Then, we performed an
asteroseismological fit  using two  independent codes:  LP-PUL and
WDEC.  From  the fits with LP-PUL  we obtained solutions
characterized   by  high   stellar  mass  of   $\sim  0.878M_{\sun}$, 
15-20\% higher than the spectroscopic value, and $T_{\rm   eff}$ 
around $11\, 200$ and $11\, 600$ K.  The best fit model
obtained with WDEC also shows a high mass of $0.815 M_{\sun}$ and an
effective temperature of $11\, 000$ K. The high mass solutions are expected given the large number of 
periods and the period spacing required to fit all modes simultaneously, since the period spacing 
decreases when mass increases and thus there are more theoretical modes in a given period range. 
Finally, all possible solutions are
characterized  by thick  hydrogen envelopes.

\subsubsection{Atmospheric parameters of GD 1212}
\label{discusion}

From the seismological study of GD 1212 using an improved list of
observed mode we obtained a best fit model characterized by $M_{\star} =0.632 M_{\sun}$ 
and $T_{\rm eff} = 10\, 737$ K. The asteroseismic
stellar mass is somewhat higher than the spectroscopic determinations
from \citet{2011ApJ...743..138G} with the 3D convection corrections
from \citet{2013A&A...559A.104T}, set at $0.619\pm 0.027
M_{\sun}$. On the other hand, from our asteroseismological study of
GD 1212 considering the period list from \citet{2014ApJ...789...85H}
we obtained solutions characterized with a high stellar mass. Using
the model grid computed with LPCODE we obtained an stellar mass $\sim
0.88 M_{\sun}$. Considering the asymptotic period spacing estimated
by \citet{2014ApJ...789...85H} of $\Delta \Pi = 41.5 \pm 2.5$ s and
the spectroscopic effective temperature $10\, 970 \pm 170$ K the
stellar mass drops to $0.770 \pm 0.067 M_{\sun}$. Also, using the
WDEC model grid, we also obtained a high mass solution, with a stellar
mass of $0.815M_{\sun}$.   The process of extracting the pulsation
periods for GD 1212, and perhaps for the cool ZZ Ceti stars showing a
rich pulsation spectra, appears to be somewhat dependent of the
reduction process \citep{2017arXiv170907004H}. Then, we must search for other independent data to
uncover the most compatible period spectra and thus seismological
solution. To this end, we search for spectroscopic and
photometric determinations of the effective temperature and surface
gravity in the literature. We used observed spectra taken by other
authors and re-determine the atmospheric parameters using up-to-date
atmosphere models. Our results are listed in table
\ref{gd1212-det}. In this table, determinations of the atmospheric
parameters using spectroscopy are in rows 1  to 7,  while  rows   8
to  11 correspond  to determinations  based on photometric  data (see
Table \ref{foto}) and parallax from  \citet{2009AJ....137.4547S}. 
We also determined the stellar mass
using our white dwarf cooling models. Finally, we include the
determinations with and without applying the 3D convection correction
for the spectroscopic determinations. 

\begin{table*}
  \caption{Determination  of   GD  1212  atmosphere   parameters  from
    different authors. Rows 1  to 7 correspond to determinations based
    on  spectroscopic   data,  while  rows   8  to  11   correspond  to
    determinations  based on photometric  data (see  Table \ref{foto})
    and             parallax            determinations            from
    \citet{2009AJ....137.4547S}.          \\         Notes:         1-
    \citet{2011ApJ...743..138G}       using      spectroscopy.      2-
    \citet{2017arXiv170907004H} using spectroscopy  3-  
   \citet{2004AJ....127.1702K}       using      spectroscopy.      4-
    \citet{2007ApJ...654..499K},             spectrum             from
    \citet{2004AJ....127.1702K}.            5-Spectrum            from
    \citet{2004AJ....127.1702K}     fitted     with    models     from
    \citet{2012A&A...538A..13K}.        6-        Spectrum        from
    \citet{2004AJ....127.1702K}     fitted     with    models     from
    \citet{2010MmSAI..81..921K}.        7-        Spectrum        from
    \citet{2011ApJ...743..138G}     fitted     with    models     from
    \citet{2010MmSAI..81..921K}.    8-    Photometric   result    from
    \citet{2012ApJS..199...29G}. 9- Photometric  data from SDSS, GALEX
    and    2MASS    and    parallax    fitted   with    models    from
    \citet{2012A&A...538A..13K}.  10- Photometric  data  from SDSS  and
    GALEX     and     parallax     fitted     with     models     from
    \citet{2010MmSAI..81..921K}.  11- Photometric  data  BVIJHK colors
    and    GALEX    and    parallax    fitted   with    models    from
    \citet{2010MmSAI..81..921K}.}
\begin{tabular}{cccccccc}
\hline\hline
  & Ref. & $T_{\rm eff}$ [K] & $\log g$ & $M_{\star}/M_{\sun}$ & $T_{\rm eff}$ [K] & $\log g$ & $M_{\star}/M_{\sun}$\\
  &      & \multicolumn{3}{c}{non - 3D} & \multicolumn{3}{c}{3D - corrected}  \\
\hline
1 & \citet{2011ApJ...743..138G} &  $11\, 270 \pm 165$ & $8.18 \pm 0.05$ & $0.705\pm 0.040$ & $10\, 970\pm 170$ & $8.03\pm 0.05$ & $0.619\pm 0.027$\\
2 & \citet{2017arXiv170907004H} & $ 11\, 280 \pm 140$ & $8.144 \pm 0.040$ & $0.684\pm 0.023$ & $10\, 980\pm 140$ & $7.995\pm 0.04$ & $0.600\pm 0.021$\\
3 & \citet{2004AJ....127.1702K} &  $10\, 960\pm 75$  & $8.20\pm 0.10$ & $0.714\pm 0.087$ & $11\, 012\pm 75$ & $7.98\pm 0.10$ & $0.592\pm 0.075$ \\
4 & \citet{2007ApJ...654..499K} &  $11\, 010\pm 210$ & $8.05\pm 0.15$ & $0.630\pm 0.100$ & $11\, 093\pm 210$ & $7.85\pm 0.15$ & $0.526\pm 0.093$\\
5 & This paper &  $11\,130\pm 200$ & $8.12\pm 0.10$ & $0.669\pm 0.078$ & $11\, 228\pm 200$ & $7.92\pm 0.10$ & $0.561\pm 0.065$ \\
6 & This paper & $11\, 770\pm 75$ & $8.27\pm 0.05$ & $0.764\pm 0.048$ & $11\, 445\pm 103$ & $8.17\pm 0.07$ & $0.698\pm 0.062$\\
7 & This paper & $11\, 573\pm 23$ & $8.04\pm 0.01$ & $0.627\pm 0.009$ & $11\, 251\pm 33$ & $7.94\pm 0.02$ & $0.573\pm 0.014$\\
\hline
8 & \citet{2012ApJS..199...29G} &  $10\, 940\pm 320$ & $8.25\pm 0.03$ & $0.747\pm 0.023$ & $\cdots$   & $\cdots$ & $\cdots$     \\
9 & This paper  & $10\, 860\pm 30$ & $8.25\pm 0.02$ & $0.747\pm 0.022$ & $\cdots$ & $\cdots$ & $\cdots$        \\
10 & This paper  & $10\, 963\pm 114$ & $8.23\pm 0.04$ & $ 0.734\pm 0.039 $ & $\cdots$  & $\cdots$ & $\cdots$  \\
11 & This paper & $11\, 153\pm 193$ & $8.28\pm 0.21$ & $ 0.771\pm 0.182 $ & $\cdots$  & $\cdots$ & $\cdots$  \\
\hline\hline
\end{tabular}
\label{gd1212-det}
\end{table*}

\begin{figure}
\begin{center}
\includegraphics[clip,width=0.5\textwidth]{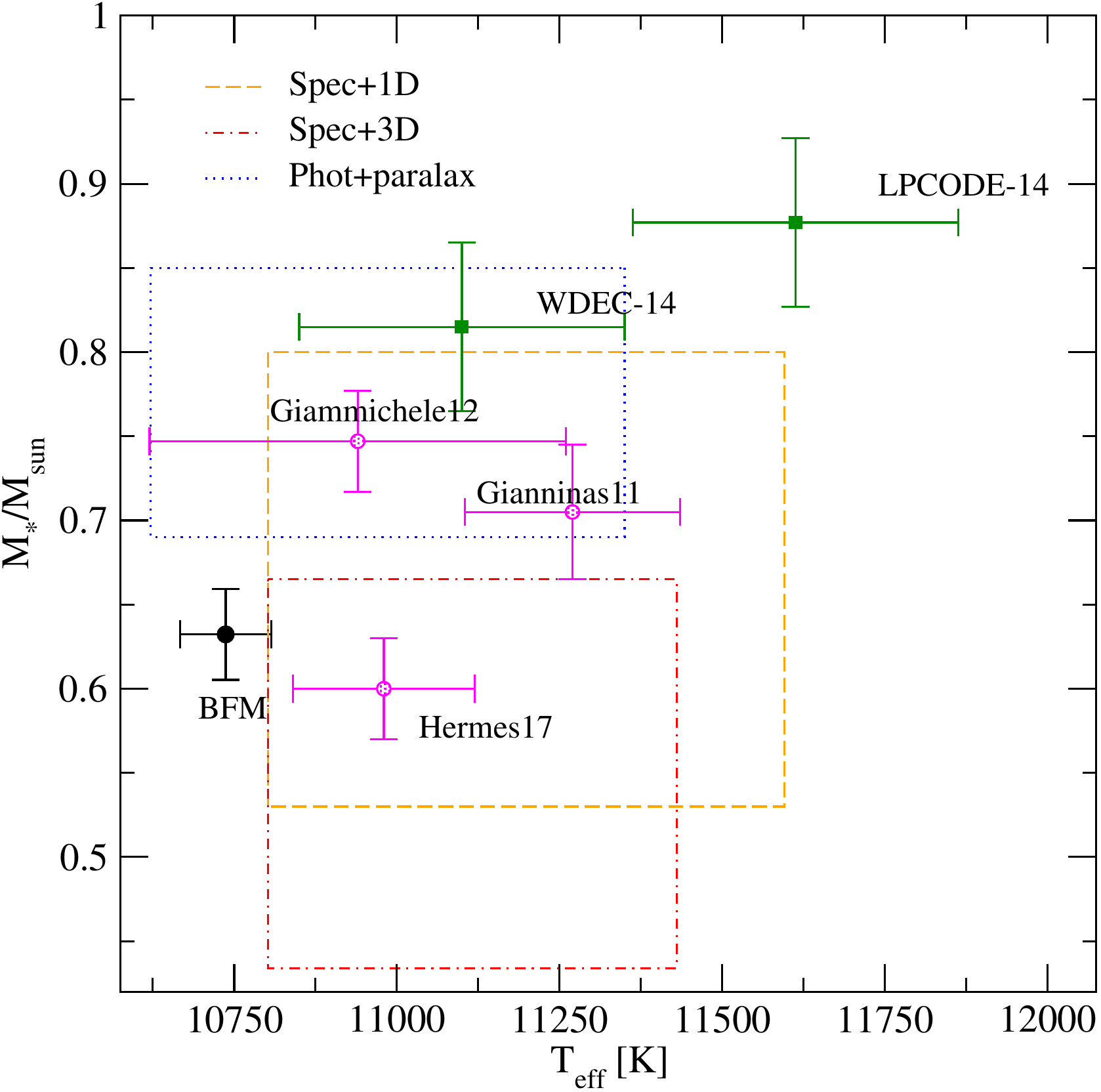}
\caption{Determinations of the effective temperature and stellar mass
  for GD 1212. The boxes correspond to the parameter range from the
  different determinations using spectroscopy,  with (Spec+3D) and
  without (Spec+1D) the 3D convection correction, and photometry
  combined with the parallax (Phot+parallax). Determinations from
  \citet{2011ApJ...743..138G}, \citet{2014ApJ...789...85H} and
  \citet{2012ApJS..199...29G} are plotted as references as hollow
  circles. The solid black circle represents the position of the best fit
  model obtained in this work. Solid squares corresponds to the
  seismological solutions  using the period list from
  \citet{2014ApJ...789...85H} obtained using the model grid computed with LPCODE (LPCODE-14) and WDEC (WDEC-14).}
\label{GD1212-todo}
\end{center}
\end{figure}

We compare  the determinations  of the  effective temperature  and the
stellar  mass for  GD 1212  using the  different techniques  discussed
above.  The results  are summarized  in Figure  \ref{GD1212-todo}. The
boxes   correspond  to   the  parameter   range  from   the  different
determinations using spectroscopy, with  and without the 3D convection
correction, and  photometry (see references  in the figure).  Our best
fit  model  is  depicted  by  a  solid  circle,  while  the  solutions
corresponding to  the asteroseismological  fits using the  period list
from \citet{2014ApJ...789...85H}  are depicted  as solid  squares. Our
best  fit   model  is  in   good  agreement  with   the  spectroscopic
determinations within the uncertainties.  The stellar  mass is  somewhat lower  than that  from
photometric  determinations  but  the   effective  temperature  is  in
excellent agreement, and consistent with a  cool ZZ Ceti star. Then we
conclude that the list of periods  shown in the right columns of table
\ref{GD1212-obs} are compatible with the photometric and spectroscopic
determinations and  is most likely to  be the the real  period spectra.

\begin{table}
\caption{Photometric data for GD 1212.}
\begin{tabular}{cccc}
\hline\hline
& mag & err & source\\
\hline 
u & 13.653 & 0.039 & SDSS\\
g & 13.267 & 0.200 &  SDSS\\
r & 13.374 & 0.018 &  SDSS\\
i & 13.547 & 0.018 &  SDSS\\
z & 13.766 & 0.021 &  SDSS\\
\hline
B & 13.440 & 0.061 &  \citet{2002ApJ...571..512H} \\
V & 13.260 & 0.048 &  \citet{2002ApJ...571..512H}\\
I & 13.240 & 0.028 & \citet{2009AJ....137.4547S} \\
J & 13.339 & 0.029 & \citet{2003yCat.2246....0C} \\
H & 13.341 & 0.023 &  \citet{2003yCat.2246....0C}\\
K & 13.35  & 0.031 &  \citet{2003yCat.2246....0C} \\
\hline
FUV & 15.714 & 0.150 & GALEX \\
NUV & 14.228 & 0.182 & GALEX \\
\hline\hline
parallax (mas) & 62.7 & 1.7 & \citet{2009AJ....137.4547S} \\
\hline
\label{foto}
\end{tabular}
\end{table}

\section{Summary and conclusions}
\label{conclusions}

In this  paper we  have presented an asteroseismological study   of
the first four published ZZ Ceti stars observed with the  \emph{Kepler}
spacecraft. We have employed an updated version of the grid of fully
evolutionary models presented in
\citet{2012MNRAS.420.1462R,2013ApJ...779...58R}. In our seismological
analysis, along with the period list, we consider additional
information coming from the detection of  rotational frequency
splittings or sequences of possible consecutive radial order modes,
i.e., period spacing value.     We summarize our results below:

\begin{itemize}
  
\item{}For KIC  11911480, we  find  a seismological  mass in  good agreement
with  the  spectroscopic  mass. Regarding  the  effective temperature,
we  find   a  higher   value  from   seismology  than spectroscopy. It
is important to note that the atmospheric parameters determined from
spectroscopy and asteroseismology can differ beyond the systematic
uncertainties, since spectroscopy is measuring the top of the
atmosphere and asteroseismology is probing the base of the convection
zone. In particular, the effective temperature characterizing our
seismological models is related to the luminosity and radius of the
model, while that from spectroscopy can vary from layer to
layer. Also, using the rotation coefficients and the frequency
spacings found by \citet{2014MNRAS.438.3086G} for three identified
dipole modes, we obtained a rotation period of $3.36 \pm 0.2$ days.

\item{} In  the case  of J113655.17+040952.6,  we found a  seismological  mass of
$0.570  M_{\sun}$ and effective temperature of $12\, 060$ K. The seismological mass is lower than that from spectroscopy but in 
agreement within the uncertainties. The seismological effective temperature is $\sim
300$ K lower than the spectroscopic value from \citet{2011ApJ...743..138G} with 3D correction
but in excellent agreement with that using  \citet{2010MmSAI..81..921K} atmosphere
models.  Finally, we determine a rotation period of 2.6 d from the
frequency spacings for the three $\ell =1$ modes identified by
\citet{2015MNRAS.451.1701H} and the rotational coefficients
corresponding to our best fit model.

\item{} KIC 4552982 is a red--edge ZZ Ceti with 18 detected periods. In this
case we found a seismological solution with a stellar mass of
0.745$M_{\sun}$ and effective temperature $11\, 110$ K, compatible
with spectroscopic determinations. The asymptotic period spacing for
dipole modes for our seismological solution (50.50 s) seems long as
compared to the period spacing estimated by
\citet{2015ApJ...809...14B} (41.9 s). However the forward period
spacing itself is compatible with the observations, as shown in figure
\ref{DP-KIC}, since the asymptotic regime is reached for periods
longer than 2000 s.  Finally, our best fit model is characterized by a
very thin hydrogen envelope mass, which could be related to the
outburst nature reported by \citet{2015ApJ...809...14B}. Whether this
is a common characteristic between all the outburst ZZ Cetis or not is
beyond the scope of this work and will be studied in a future paper. 

\item{} Finally,  GD  1212 is also a red--edge ZZ Ceti with 9 independent
pulsation periods. We obtained a best fit model characterized by $M_{\star}
= 0.632 M_{\sun}$ and $T_{\rm eff}=10\, 922$ K. The stellar mass is
somewhat higher than the spectroscopic value, but the effective
temperature is in excellent agreement. We also fit the period list
reported in \citet{2014ApJ...789...85H} and obtained a high stellar
mass solution ($\sim 0.88M_{\sun}$). However, other determinations of
the atmospheric parameters from photometry combined with parallax and
spectroscopy point to a lower value of the stellar mass, closer to
$M_{\star}= 0.66 M_{\sun}$, and thus compatible with the seismological
solution for the update period list of GD 1212 presented in this work.

\end{itemize}

On the basis of the recent study by De Ger\'onimo et al. (2017b, submitted), 
we can assume that the uncertainties in stellar mass, effective temperature and thickness of the H-rich envelope of our
asteroseismological models due to the uncertainties in the prior
evolution of the WD progenitor stars, as the TP-AGB, amount to 
$\Delta M_{\star}/M_{\star} \lesssim 0.05$,
$\Delta T_{\rm eff} \lesssim 300$ K and a factor of two, respectively. We empasize that these
uncertainties are more realistic than the formal errors quoted in the
Tables of this paper that correspond to the internal uncertainties due
to the period-fit procedure.

Note that, generally speaking, asteroseismology of the stars observed
by {\it Kepler} can be analyzed in the same way as the ones with just
ground base observations. At the hot end, ZZ Ceti stars shows short
periods with low radial order, that propagates in the inner region of
the star, giving more information about its internal structure. Also,
it appears to be no additional "noise" in the period list
determinations due to pointing corrections of the {\it Kepler}
spacecraft, as can be seen by comparing the asteroseismological
analysis for KIC 11911480 and J3611$+$0409. 

For cool ZZ Cetis, we see a rich period spectra, with mostly long
periods with high radial order. In this case, more periods does not
mean more information, since high radial order modes propagates in the
outer region of the star. However, we can extract an additional
parameter from the period spectra: the mean period spacing. This is
particularly the case for KIC 452982, giving the chance to estimate
the stellar mass somewhat independently form the period-to-period
fit. In addition, we use the spectroscopic parameters as a restriction
to the best fit model.  For GD 1212, the reduction process involving
the extraction of the period list from the light curve is quite
problematic. Thus we needed the help of photometry and spectroscopy to
select the most probable period spectra for GD 1212.  

Together with the studies of
\citet{2012MNRAS.420.1462R,2013ApJ...779...58R} for an  ensemble of
ZZ Ceti stars observed from the ground, the results for ZZ Cetis
scrutinized with the \emph{Kepler} mission  from space  presented in
this work  complete the first thorough asteroseismological survey of
pulsating DA WDs based on fully  evolutionary pulsation models. We are
planning to expand this survey by performing new asteroseismological
analysis of a larger number of DAV stars, including the new ZZ Ceti stars
observed with the{\emph Kepler} spacecraft and also from the SDSS.

\begin{acknowledgements}
We wish to thank our anonymous referee for the constructive comments
and suggestions.   
 Partial financial support from
this research comes from CNPq and PRONEX-FAPERGS/CNPq
(Brazil).
Part of this work was supported by AGENCIA through the Programa de
Modernizaci\'on Tecnol\'gica BID 1728/OC-AR and the PIP
112-200801-00940 grant from CONICET. D.K. received support from programme Science without
Borders, MCIT/MEC-Brazil. JH received support from NASA through Hubble Fellowship grant \#HST-HF2-51357.001-A, awarded by the Space Telescope Science Institute, which is operated by the Association of Universities for Research in Astronomy, Incorporated, under NASA contract NAS5-26555. AK acknowledges support from the Czech
Science Foundation (15-15943S). AG gratefully  acknowledges the
support of the NSF under grant AST-1312678, and NASA under grant
NNX14AF65G.  We thank Ingrid Pelisoli, Gustavo Ourique and Stephane
Vennes for useful discussions.  This research has made use of NASA
Astrophysics Data System. 
\end{acknowledgements}

\bibliographystyle{aasjournal} 
\bibliography{Romero-apj} 






\end{document}